%% file: DSAMARL.tex
\newif\ifOneCol

\ifOneCol
    \documentclass[12pt,onecolumn,draftclsnofoot]{IEEEtran}
    \renewcommand{\baselinestretch}{1.425}
\else
    \documentclass[journal,10pt,twocolumn]{IEEEtran}
    \renewcommand{\baselinestretch}{1.0}
\fi
\newcommand{\Figref}[1]{Figure~\ref{fig:#1}}
\usepackage{graphicx}
\usepackage{epsfig}
\usepackage{subfigure}
\usepackage{amssymb}
\usepackage{amsbsy}
\usepackage{amsmath}
\usepackage{steinmetz}
\usepackage{tabularx}
\usepackage[T1]{fontenc}
\usepackage[font=scriptsize]{caption}
\usepackage{cite}
\usepackage{url}

\usepackage{amsfonts}
  
\usepackage{color}
\usepackage{float}
\usepackage{times,amsmath,epsfig}
\usepackage{xspace,latexsym,syntonly}
\usepackage{amssymb}
\usepackage{textcomp}
\usepackage{bbm}
\usepackage{xcolor}

\usepackage{algorithm}

\usepackage[noend]{algpseudocode}
\algnewcommand\INPUT{\item[\textbf{Input:}]}%
\algnewcommand\OUTPUT{\item[\textbf{Output:}]}%

\newtheorem{definition}{Definition}
\newtheorem{theorem}{Theorem}

\newtheorem{lemma}{Lemma}

\input{macros}

\begin{document}\title{SINR-Aware Deep Reinforcement Learning for Distributed Dynamic Channel Allocation in Cognitive Interference Networks}
\author{{Yaniv Cohen, Tomer Gafni, Ronen Greenberg, Kobi Cohen (\emph{Senior Member, IEEE})}
\thanks{Corresponding author: K. Cohen, yakovsec@bgu.ac.il. K. Cohen and T. Gafni are with the School of Electrical and Computer Engineering, Ben-Gurion University of the Negev, Beer Sheva 8410501 Israel. Y. Cohen and R. Greenberg are with Elbit Radio Ltd., Israel.\\
This work was supported by the Israel Ministry of Economy under the Magnet consortium program.\\
This work has been submitted to the IEEE for possible publication. Copyright may be transferred without notice, after which this version may no longer be accessible.
}}
\date{}
\maketitle
\begin{abstract}
We consider the problem of dynamic channel allocation (DCA) in cognitive communication networks with the goal of maximizing a global signal-to-interference-plus-noise ratio (SINR) measure under a specified target quality of service (QoS)-SINR for each network. The shared bandwidth is partitioned into $K$ channels with frequency separation. In contrast to the majority of existing studies that assume perfect orthogonality or a one-to-one user-channel allocation mapping, this paper focuses on real-world systems experiencing inter-carrier interference (ICI) and channel reuse by multiple large-scale networks. This realistic scenario significantly increases the problem dimension, rendering existing algorithms inefficient. We propose a novel multi-agent reinforcement learning (RL) framework for distributed DCA, named Channel Allocation RL To Overlapped Networks (CARLTON). The CARLTON framework is based on the Centralized Training with Decentralized Execution (CTDE) paradigm, utilizing the DeepMellow value-based RL algorithm. To ensure robust performance in the interference-laden environment we address, CARLTON employs a low-dimensional representation of observations, generating a QoS-type measure while maximizing a global SINR measure and ensuring the target QoS-SINR for each network. Our results demonstrate exceptional performance and robust generalization, showcasing superior efficiency compared to alternative state-of-the-art methods, while achieving a marginally diminished performance relative to a fully centralized approach.
\end{abstract}

%

\section{Introduction}
\label{sec:introduction}

The growing need for wireless communications, alongside the challenges posed by spectrum scarcity, has prompted the advancement of effective Dynamic Channel Allocation (DCA) schemes for emerging cognitive communication network technologies. Our focus is on DCA within large-scale networks, aiming to optimize the allocation of frequency channels among networks sharing the same frequency band. Unlike static channel allocation, where frequencies remain fixed, DCA adapts to changing network conditions by dynamically allocating channels based on factors such as interference and signal strength, impacting user rates in the network. DCA enhances the efficiency and flexibility of wireless networks, ensuring efficient channel utilization and mitigating the impact of interference, leading to improved overall performance and quality of service (QoS) . Recent studies on this topic, which involves learning the environment and channel allocation algorithms, have employed frameworks such as multi-armed bandits (MAB) \cite{liu2012cooperative, Tekin_2012_Online, liu2012learning, nayyar2016regret, avner2016multi, bistritz2018distributed, gafni2020learning, wang2021decentralized, gafni2022learning}, stable matching \cite{leshem2012multichannel, gafni2022distributed}, game theoretic optimization and congestion control \cite{han2005fair, menache2008rate, menache2011network, liu2012cooperative, law2012price, cohen2013game, wu2013fasa, singh2016combined, cohen2016distributedToN, cohen2017distributed, malachi2020queue, 
wang2021decentralized}, and, more recently, deep reinforcement learning (DRL) for multi-user scenarios\cite{naparstek2017deep, naparstek2018deep, chang2018distributive, yang2018dynamic, su2019reinforcement, zhang2019energy, lee2020dqn,
ning2020reinforcement, xu2020intelligent,
obite2021overview, du2022multi, adeogun2022multi, bokobza2023deep, adeogun2023distributed, paul2023multi}. The latter was initially explored in our earlier work (Naparstek and Cohen \cite{naparstek2017deep, naparstek2018deep}) within a multi-agent framework, following single-agent DRL research in \cite{wang2018deep}, paving the way for a significant amount of subsequent research in the wireless communications community.

\subsection{DRL Algorithms for DCA}

DCA algorithms can be classified into two distinct branches: Centralized and distributed methods. Related studies within the centralized approach have utilized techniques such as graph vertex coloring, where adjacent vertices (representing users or local networks) are modeled with strong mutual interference, necessitating their allocation to different channels. Colors represent orthogonal channels, effectively coloring the graph. This paper addresses more challenging scenarios where interference exists between channels, and varies across the graph, for instance, decreasing with geographical distance, contrasting the binary interference assumption in graph coloring models. Additionally, centralized DRL methods have been recently proposed (see our recent work \cite{paul2023multi} and references therein). The limitation of the centralized approach lies in the communication overhead and computational complexity required for solving centralized DCA optimization. Consequently, in many applications, the preference is for distributed DCA, as it is widely considered more suitable for various wireless communication scenarios due to the intricate nature of the problem. 

In this paper, we focus on DRL methodology to address the distributed DCA problem. In recent years, the notable achievements of DRL in attaining strong performance surpassing human capabilities across various game domains have captured the attention of researchers from various disciplines. Prominent examples include its successes in games such as Atari \cite{mnih2013playing, hessel2018rainbow}, Go \cite{schrittwieser2020mastering}, and even in complicated game such as StarCraft II \cite{vinyals2019grandmaster}. This success has motivated researchers to explore the development of DRL in diverse fields such as robotics,  biology, chemistry, finance \cite{li2017deep}, as well as wireless communications  \cite{xing2006dynamic,
su2019reinforcement,
yang2018dynamic,
lee2020dqn,
chang2018distributive,
bokobza2023deep,
naparstek2018deep,
naparstek2017deep,
zhang2019energy,
ning2020reinforcement,
obite2021overview,
xu2020intelligent}. This approach is particularly promising in the topic of DCA, where the collection of data becomes inefficient due to the high dimensionality of the state space resulting from the numerous uncertainties inherent in the environment. Particularly, recent study \cite{adeogun2022multi} proposed a model-free Multi-Agent Reinforcement Learning (MARL) independent Q-learning (IQL) table approach to solve the distributed DCA problem over orthogonal channels. While this work demonstrates the potential of a simple RL technique for addressing the DCA challenge, it is limited to small-scale systems due to the high dimensionality of the state space in the DCA problem. To overcome this limitation, the state space was reduced and strongly quantized in the conducted study. Additionally, the IQL technique, both in a general context and specifically in the domain of radio systems, imposes constraints on system scalability. This arises from the inherent characteristic wherein each agent acquires an independent policy, which is greatly dependent on the policies formulated by the other participating agents. Furthermore, recently in \cite{adeogun2023distributed}, the authors employed the Double Deep Q-network (DDQN) \cite{van2016deep} as the RL optimization mechanism to solve DCA in multi-agent systems. This utilization of the DDQN facilitates the exploration of more effective strategies. The transition from a tabular approach to neural networks (NN) diminishes the dependence on quantization techniques employed in previous research, leading to an improved generalization of the problem. In \cite{du2022multi}, an attention-based Graph Neural Network (GNN) architecture known as GA-Net, was introduced. The study addressed DCA in 6G sub-networks using a centralized training approach and a decentralized execution paradigm. The authors assumed channel orthogonality and employed a modified Soft-Actor-Critic (SAC) algorithm as the optimization technique for MARL system. By combining the hard attention layer with the Graph Attention Network (GAT), the proposed algorithm demonstrated improved performance by effectively extracting relevant features that capture the relationships between different sub-networks compared to other techniques examined in \cite{du2022multi}. Motivated by the recent success of DRL in DCA, our paper takes on a challenge that has not yet been effectively addressed. Specifically, here we focus on developing a DRL framework for distributed DCA among large-scale cognitive interference networks. This framework accommodates channel reuse, continuous interference over the network graph, all without assuming complete orthogonality due to ICI.

\subsection{Main Results}

Multi-agent DRL for DCA has garnered significant attention in recent years, spurred by our earlier work \cite{naparstek2017deep, naparstek2018deep}. In contrast to related recent studies that predominantly centered around relatively small-scale networks, often addressing singular sensor-controller low-power links within enclosed environments (e.g., robotic systems), the present study introduces a distributed DCA algorithm tailored for large-scale network configurations. This algorithm places paramount importance on attaining elevated channel quality (CQ) while concurrently minimizing convergence time in outdoor scenarios. Additionally, in contradistinction to existing studies that assumed orthogonal channel conditions, our approach confronts the intricate complications arising from ICI across channels, where orthogonality is not preserved. It underscores the need for minimal spectrum mobility, an imperative requirement for distributed networks supporting a multitude of users. The realistic challenging scenario considered in this paper significantly increases the search space, rendering existing algorithms inefficient.

To tackle this problem, we propose a novel MARL framework for distributed DCA, named Channel Allocation RL To Overlapped Networks (CARLTON). The CARLTON framework is based on the Centralized Training with Decentralized Execution (CTDE) paradigm, utilizing the DeepMellow value-based RL algorithm \cite{kim2019deepmellow}. DeepMellow's design mitigates the necessity for a target network in comparison to conventional DQN algorithms. To ensure robust performance in the interference-laden environment we address, CARLTON quantizes observations to generate a QoS-type measure while maximizing a global SINR measure and guaranteeing the target QoS-SINR for each network. CARLTON incorporates a novel reward framework that trades off personal rewards with social rewards, encouraging cooperative behavior among distinct networks incapable of information exchange. Moreover, we introduce a preprocessing mechanism designed to enhance the algorithm's adaptability to diverse network sizes. Within our specific problem domain, each network receives an exclusive sensed vector, representing only a fraction of the comprehensive state information inherent in the MARL environment. Consequently, the problem discussed herein inherently falls within the domain of Partially Observable Markov Decision Process (POMDP) problems. 

The proposed CARLTON algorithm has demonstrated exceptional performance and robust generalization capabilities, outperforming various alternative methods. This includes surpassing the common random agent (RA) benchmark \cite{cohen2017distributed} and, notably, outperforming the state-of-the-art method for our model, the well-known Jamming Avoidance Response (JAR) algorithm \cite{huang2022prospects}. CARLTON demonstrated a remarkable approximate margin of superiority of $45\%$ over the RA benchmark and $20\%$ over the JAR algorithm, respectively. Furthermore, it exhibits only a slight discrepancy of approximately $2.5\%$ in performance compared to a centralized approach computed by graph coloring. This level of effectiveness positions CARLTON as an outstanding algorithm for efficiently managing DCA among distributed cognitive networks in complex interference environments.

\section{System Model and Problem Formulation}\label{PD}
\label{sec:system}

\subsection{Description of the System}

\begin{figure}[ht!]
\label{fig:channels}
\vspace{-0.0cm}
\begin{center}
\advance\leftskip-2cm
\advance\rightskip-3cm
\includegraphics[width=0.56 \textwidth,trim={0.cm 0.0cm 0.0cm 0.0cm},clip ]{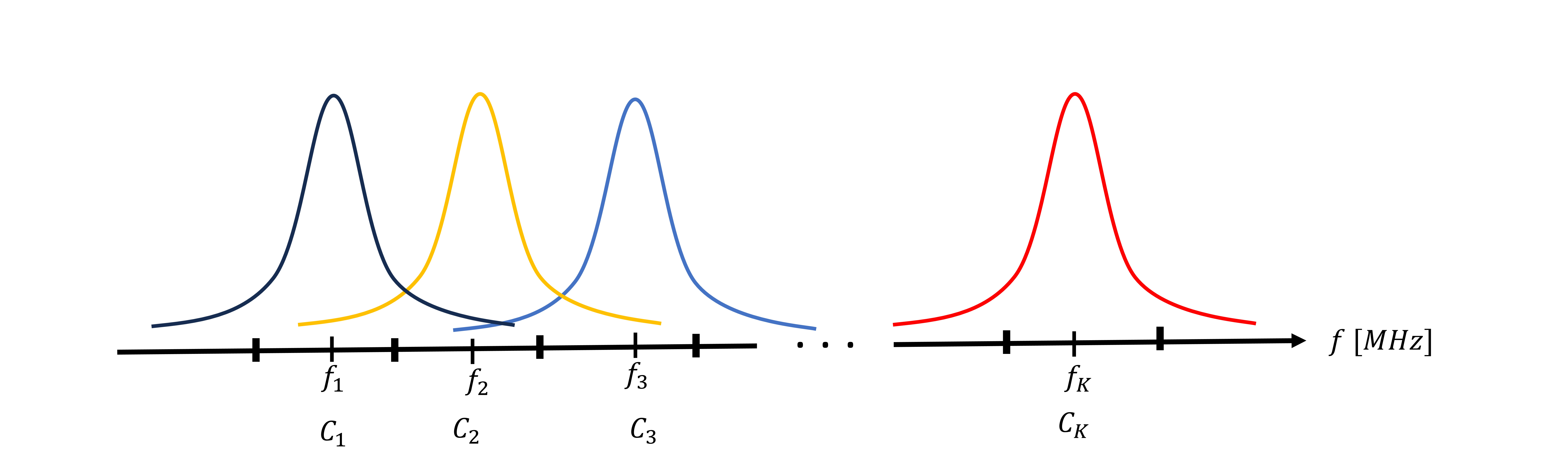} 
\caption[]{An illustration of the non-orthogonal channel partition. Channel $i$, denoted by $C_i$ in the figure, refers to the carrier frequency $f_i$.}
\label{fig:1}
\end{center}
\end{figure}

We consider a system comprising $N$ wireless networks represented by the set $\mathcal{N} = \{\mathcal{N}^1, ..., \mathcal{N}^N \}$. Each network (say $\mathcal{N}^n$) accommodates $M^{n}$ users, with $M^n\in\{1, 2, ..., M_{max}\}$, and $M_{max}$ denoting the maximum number of users among all networks. A total bandwidth (TB) is equally divided into $K$ equal-sized overlapping channels with bandwidth $B$ denoted by the set $\mathcal{K} = \{1,2, ..., K\}$ such that each network can operate on a single channel at each time slot. Orthogonality is not maintained across carriers, resulting in ICI between channels. In this context, we index the channels based on their serial numbers in $\mathcal{K}$, as depicted in \Figref{1}.

To model the interference between networks, we employ the well-established Egli model \cite{egli1957radio}. The Egli model, developed for UHF and VHF outdoor signals, is commonly used for modeling real-world radio frequency propagation. Based on the model, the SINR of each user at each available channel is computed. The SINR accounts for the interference from the other networks in the environment, and the additive thermal noise at the receiver. In what follows, to describe the computation, we use the indexing notations
\begin{equation}
(\cdot)^{l,n}_{m,j,k}    
\end{equation}
to represent a parameter regarding a pair of users, denoted as user $m \in \mathcal{N}^l$ and user $j \in \mathcal{N}^n$. The SINR is computed with respect to channel $k$. When the parameter pertains to a single user, irrespective of the presence of a second user, we specify only a single user. Specifically, the SINR of users within network $\mathcal{N}^{n} \in \mathcal{N}$ on channel $k\in \mathcal{K}$ is computed as follows. Consider users $i, j\in \mathcal{N}^n$, where user $i$ transmits a message to user $j$, operating on channel $k\in \mathcal{K}$. The received $\text{SINR}$ at the receiver of user $j$ is given by:
\begin{equation}
\text{SINR}^{n,n}_{i,j,k} = \frac{\text{PR}^{n,n}_{i,j,k}}{I_T + I^{n}_{j,k}},
\end{equation}  
where $\text{PR}^{n,n}_{i,j,k}$ is the signal power at the receiver of user $j\in \mathcal{N}^n$ transmitted by user $i\in \mathcal{N}^n$ on channel $k$, $I_T$ is the additive thermal noise power, and $I^{n}_{j,k}$ is the interference power at the receiver of user $j\in \mathcal{N}^n$ on channel $k$. This interference is attributed to all other networks $\mathcal{N}\setminus\mathcal{N}^n$ transmitting on any possible channel, with ICI modeled by the Egli model, as computed later. Interference between users within the same network is avoided by the Medium Access Control (MAC) scheduler mechanism. The computation of the received power is given by:
\begin{equation}\label{PR}
\text{PR}^{n,n}_{i,j,k}= \text{PT}^{n}_{j,k} - \text{PL}^{n,n}_{i,j,k},
\end{equation}
where $\text{PT}^{n}_{j,k}$ is the transmitter power of user $j\in\mathcal{N}^n$ on channel $k$ (possibly broadcasted signal irrespective of user $j \in \mathcal{N}^n$) in dBW, and $\text{PL}^{n,n}_{i,j,k}$ is the path loss between user $j\in\mathcal{N}^n$ and user $i\in\mathcal{N}^n$ on channel $k$, which obeys the Egli model and is given by:
\begin{equation}
\begin{split}
\text{PL}^{n,n}_{i,j,k}= & \ 40\log_{10}(d^{n,n}_{i,j}) \\ & - 20\log_{10}\left( \frac{40}{f_{k}}\right) - 20\log_{10}(H^{n}_{T_{i}} H^{n}_{R_{j}}) \\ & - 10\log_{10}(G^{n}_{T_{i}} G^{n}_{R_{j}}) \ [dB],
\end{split}
\end{equation}
where $d^{n,n}_{i,j}$ is the euclidean distance between users $j\in\mathcal{N}^n$ and $i\in\mathcal{N}^n$ $i$ in [meters], $f_{k}$ is the carrier frequency of channel $k$ in [MHz], $H^{n}_{T_{j}}$ and $H^{n}_{R_i}$ are the heights of the transmit and receive antennas of users $j\in\mathcal{N}^n$ and $i\in\mathcal{N}^n$ in [meters], respectively, and $G^{n}_{T_{j}}$ and $G^{n}_{R_{i}}$ are the absolute gain of the transmitter and receiver antennas of users $j\in\mathcal{N}^n$ and $i\in\mathcal{N}^n$, respectively.

The interference power from other networks $\mathcal{N} \setminus \mathcal{N}^{n}$ at the receiver of user $j\in\mathcal{N}^{n}$ on channel $k$ is given by: 
\begin{equation}
\begin{gathered}
I^{n}_{j,k} = \sum_{l \in \{1,...,n-1,n+1,...,N\}} \Phi^{n, l}_{j, k} \\
\hspace{-7.5cm}\mbox{where}\\
 \Phi^{n, l}_{j, k} \triangleq \sum_{m \in \{1,...,M^{l}\}} \phi^{n, l}_{j, m, k},
\end{gathered}
\end{equation}
where $\phi^{n, l}_{j, m, k}$ is the interference power from user $m\in\mathcal{N}^l$, given by: 
\begin{equation}
\phi^{n, l}_{j, m, k} = \text{PR}^{n, l}_{j, m, k} - T(k,\tilde{k}) 
\end{equation} 
where $\tilde{k}$ is the channel used by network $\mathcal{N}^l$, and $T(k,\tilde{k})$ is the attenuation at the channel of interest $k$ with respect to channel $\tilde{k}$. The attenuation used in the simulations in this paper is given in Table \ref{table:1} \cite{goldsmith2005wireless}.

\begin{table}[h!]
\centering
\begin{tabular}{ |p{3.5cm}|p{3cm}|}
\hline
 Spectral Distance & Attenuation (dBm)\\
 $|k-\tilde{k}|$ &  \\
\hline 
0 & 0 \\
\hline
1 & 20 \\
\hline 
2 & 40 \\
 \hline
3& 50 \\
\hline 
4 & 60 \\
 \hline
$\geq 5$ and $|f_k - f_{\tilde{k}}| / f_k  \leq 0.05$  & 95 \\
\hline

else & 110 \\
 \hline
 
\end{tabular}
\caption{Attenuation with respect to spectral distance, where $f_k$ and $f_{\tilde{k}}$ represent the frequencies of channels $k$ and $\tilde{k}$, respectively.}
\label{table:1}
\end{table}

The thermal noise $I_T$ at the receiver is calculated as follow:
\begin{multline}
I_T = k_B \cdot \text{Temp} \cdot B \cdot \text{NF} \ [W] \ =\\ \ 10 \log_{10}(k_B \cdot \text{Temp} \cdot B) + 10\log_{10}(\text{NF})  + 30\ [dBm],
\end{multline} 
where $k_B$, $\text{Temp}$, $B$, and $\text{NF}$ stands for Boltzmann constant, temperature, bandwidth, and noise figure. For instance, at room temperature, bandwidth of $2MHz$, and $10\log_{10}(\text{NF}) = 6$, we have $I_T = -104.9 \ [dBm]$.

\subsection{Illustration of the System Environment in Real-World Simulations}
\label{environment}

To elaborate on the system model and its significant relevance in dynamic spectrum access systems, we present a comprehensive depiction of the system environment through real-world simulation settings that inspire this research. In each simulation instance, a specific number of networks, denoted as $N$, is randomly chosen. For each network, the number of users is randomly assigned, following a uniform distribution (ranging from 2 to 22 users in our simulations). The spatial distribution of users within each network is determined by a multi-variate Gaussian distribution. The expectation vector is defined by a 2D center point, and the covariance is represented by a diagonal matrix with entries $[50^2, 50^2]$ [$m^2$]. It is noteworthy that our current investigation primarily addresses a 2D spatial problem, but the methodology can be readily extended to encompass 3D dimensions. The center point for each network is determined using the procedure outlined in Algorithm \ref{alg:algo1}.

\begin{algorithm}
\caption{- Set Networks Center Points Locations}
\label{alg:algo1}
\begin{algorithmic}[1]
\INPUT $N$ \Comment{Number of networks}
\OUTPUT $\mu$  \Comment{Dictionary of all center points, $\mu \in \mathbb{R}^{2\times N}$}
\State $\mu \leftarrow \{\}$ \Comment{Define empty dictionary}

\For {$i \in \{1,..., N\}$}  
	\If {$i == 0$} 
		\State $\mu^{0}_{x} \leftarrow u_x$ s.t $U_x \sim U[-u_1 N,u_1 N]$ 
		\State $\mu^{0}_{y} \leftarrow u_y$ s.t $U_y \sim  U[-u_1 N,u_1 N]$
	\Else
		\State  $\mu^{c}$ = choose random center point from $\mu$ 
		\State r $\leftarrow$ sample from $U[x_1, x_2]$ 
		\State $\theta$ $\leftarrow$ sample from $U[0,2\pi]$
		\State $\mu^{i}_{x} \leftarrow \mu^{c}_{x} + r cos(\theta) $ 
		\State $\mu^{i}_{y} \leftarrow \mu^{c}_{y} + r sin(\theta) $ 
		
	\EndIf

	$\mu[i]$ $\leftarrow$ $(\mu^{i}_{x}, \mu^{i}_{y})$
\EndFor
\end{algorithmic}
\end{algorithm}

\begin{figure}[ht!]
\vspace{-0.0cm}
\begin{center}
\advance\leftskip-3cm
\advance\rightskip-3cm

\includegraphics[width=0.50 \textwidth,trim={0.cm 0.0cm 2cm 1cm},clip ]{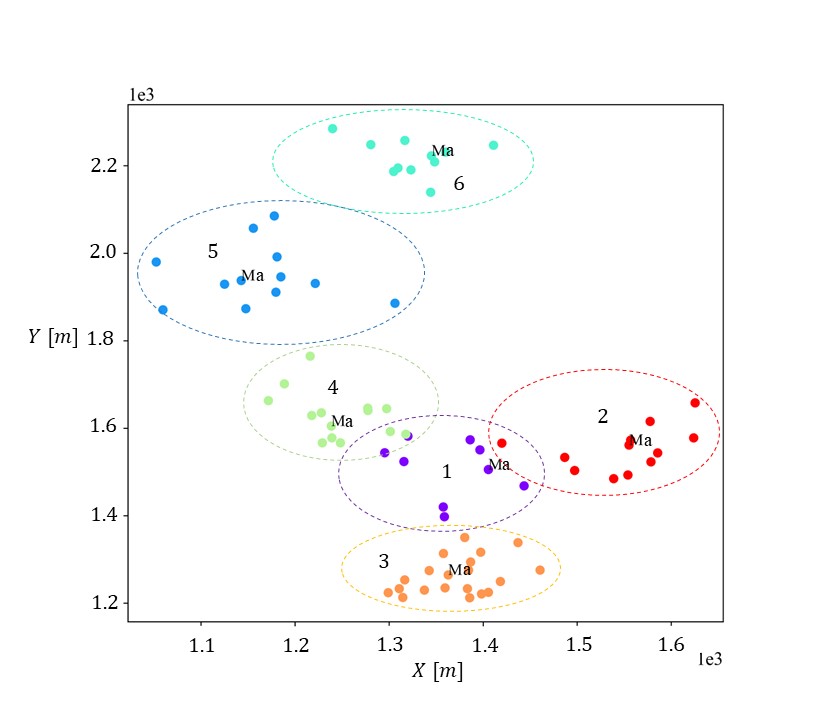} 

\caption[]{An illustration of the simulation environment involving six networks. Each color corresponds to a distinct network, distinguished by a unique serial number. The symbol 'Ma' denotes the network manager present at each respective network.}
\label{fig:2}
\end{center}
\end{figure}

In our simulation environment, we incorporate the concept that each network includes a specific user designated as the \emph{network manager}, aligning with the common practice in 5G technology where centralized units oversee local networks. The network manager is selected as the user within the network whose total Euclidean distance from all other users in the same network is minimized. This selection is based on the user's ability to maintain the shortest average distance to all other users within the given network. The role of the network manager is pivotal, taking on the responsibility of gathering information from other users within the network and disseminating decisions as feedback. This feedback mechanism fosters effective communication and coordination among users, allowing the network manager to serve as a central point for control and information exchange within its respective network, using a dedicated control channel. By fulfilling this crucial function, the network manager significantly contributes to optimizing the overall efficiency of the network's operations. \Figref{2} illustrates a representative scenario, featuring a possible configuration of networks, each accompanied by its designated network manager.

\subsection{The Objective}\label{TheObjective}

Let $c^{n}\in\{1,...,K\}$ represent the operating channel of network $\mathcal{N}^n$,and ${\text{SINR}}^{n}_{j,c^n}$ be the average SINR of user $j\in\mathcal{N}^n$, computed in consideration of all messages received from the other $M^n-1$ users in network $\mathcal{N}^n$. Then, ${\text{SINR}}^{n}_{j,c^n}$ is given by:
\begin{equation}
{\text{SINR}}^{n}_{j,c^n} =  \frac{1}{M^{n}-1}\sum_{\substack{i=1 \\ i\neq j}}^{M^{n}}  \text{SINR}^{n,n}_{j, i, c^n}.   
\end{equation}
Then, the average SINR of network $\mathcal{N}^n$, denoted by $\overline{\text{SINR}}^{n}_{c^n}$, is given by:
\begin{equation}
\overline{\text{SINR}}^{n}_{c^n} =  \frac{1}{M^{n}}\sum_{j=1}^{M^{n}}  {\text{SINR}}^{n}_{j,c^n}.
\end{equation}

The objective is to maximize the average $\overline{\text{SINR}}^{n}_{c^n}$ across all participating networks, subject to the constraint that the $\overline{\text{SINR}}^{n}_{c^n}$ for each network exceeds the target QoS-SINR required for high-quality communication, defined by SINR$^*$:
\begin{equation}
\begin{array}{l}
\displaystyle \max_{\left\{c^{n}\right\}^N} \frac{1}{N}\sum_{n=1}^{N}\overline{\text{SINR}}^{n}_{c^n}
\vspace{0.2cm}\\\displaystyle
\text{subject to }  \overline{\text{SINR}}^{n}_{c^n} \geq \text{SINR}^* \ \;\;\forall n\in \mathcal{N},
\end{array}
\end{equation}

Two crucial requirements for the algorithm are identified. Firstly, the algorithm must exhibit a short convergence time to swiftly adapt to the dynamic nature of the spectrum environment, ensuring timely and efficient allocation of available resources. Secondly, a distributed implementation is imperative to promote scalability, enhance flexibility, and reduce the computational burden on individual nodes, fostering a more resilient and adaptable system architecture.

\section{The Channel Allocation RL To Overlapped Networks (CARLTON) Algorithm}

In this section, we introduce the CARLTON algorithm designed to address the objective. We formulate the DCA problem as a MARL framework, where each network manager acts as an autonomous agent within the global framework. Consistent with standard Reinforcement Learning (RL) paradigms, the well-defined environment comprises observation, action, and reward components. Following the system model outlined in Section \ref{sec:system}, the CARLTON framework enables network managers (agents) to iteratively learn and adapt their channel allocation strategies, aiming for improved channel quality and reduced spectrum mobility until the system converges to a desirable solution. We commence by addressing and defining the mentioned requirements in detail.

\subsection{Observation Space}

The observation space encompasses a fusion of SINR measurements from all users within the network. Before the network manager makes a decision, each user transmits its individual SINR measurements for all available channels via a dedicated control channel to the network manager. This vector of SINR measurements has a dimension of $\mathbb{R}^K$ (SINR value for each channel).

The network manager, after conducting SINR sensing for all channels, receives unique SINR measurement vectors from all users in the network. These individual vectors are then concatenated to form an SINR matrix with dimensions $\mathbb{R}^{M^{n} \times K}$, where $M^{n}$ corresponds to the total number of users in network $\mathcal{N}^n$. The entry ${\text{SINR}}^{n}_{j,k}$ at row $j$ and column $k$ of the matrix, where $j=1, ..., M^{n}$ and $k=1, ..., K$, represents the SINR measurement for user $j$ on channel $k$. Additionally, we define a binary matrix $\text{BSINR}^n$ with dimensions $M^{n} \times K$, where each entry is set to either 0 or 1, indicating whether the corresponding ${\text{SINR}}^{n}_{j,k}$ value is below or above the target QoS-SINR threshold, respectively:
\begin{equation}
\text{BSINR}^{n}_{j,k} = \mathbbm{1}({\text{SINR}}^{n}_{j,k} > \text{SINR}^*),
\end{equation}
where $\mathbbm{1}$ is the indicator function, equals $1$ if ${\text{SINR}}^{n}_{j,k} > \text{SINR}^*$, or equals zero otherwise.  

Next, the network manager performs an averaging operation on the $\text{BSINR}^{n}$ matrix along the user axis to mitigate the influence of the number of users in the network. The $k$th entry of the vector for channel $k$ is given by:
\begin{equation}
\text{QV}^{n}_k = \frac{1}{M^{n}}\sum_{j=1}^{M^{n}}\text{BSINR}^{n}_{j,k}.
\end{equation}
This averaging process results in a vector representing the average number of users experiencing sufficient communication quality. We will refer to this vector as the Quality Vector ($\text{QV}^n \in \mathbb{R}^{K}$), which serves as a key observation in the simulations.

\subsection{Action Space}

The action space encompasses the set of allowable actions available to the agent at each decision point. In the current context, the action space includes all possible frequency channels and is represented by $\mathcal{K}$. Thus, at each time step, the action of network $\mathcal{N}^n$ is defined as $a^{n}(t) \in \mathcal{K}$.

\subsection{Reward}

The agents are tasked with solving a combinatorial problem, namely, establishing a one-to-one mapping between each network and a channel. The objective is to reach a state where, either upon convergence (where each network remains connected to the same chosen channel until the completion of the scenario) or at the end of the scenario, the total mean quality of the selected channels across all networks is maximized, as detailed in Section \ref{TheObjective}. 

To pursue this objective, we embrace a two-part reward value, consisting of two distinct components. The first element relates to the personal reward, denoted by $r_p$. Its magnitude is influenced by the $\text{QV}^n(t)$ value associated with the channel quality ranking selected by the agent pre-selection cycle of all other networks at time step $t$. Essentially, $r_p$ reflects a greedy mechanism aimed at selecting high-quality channels. Additionally, the personal reward experiences positive growth if the action remains unchanged. The reward computation is elucidated in Algorithm \ref{alg:algo2}.

\begin{algorithm}
\caption{CARLTON's Personal Reward, $r_p$}
\label{alg:algo2}
\begin{algorithmic}[1]
\INPUT $\text{QV}^n(t)$, $a^{n}(t)$, $a^{n}(t-1)$ \Comment{$a^{n}(t)$ is the chosen channel at time $t$}
\OUTPUT $r_p$  
\State $v \leftarrow \text{QV}_{a^n(t)}^{n}(t)$ \Comment{Extract the $\text{QV}^{n}(t)$ value of the chosen channel}

\If {$v \geq \zeta$}
	\State $r_p \leftarrow r_{desired}$
	
\Else 
	\State $\text{SQV}^{n} \leftarrow $ sort($\text{QV}^n(t)$)
	\State $i \leftarrow$ 0
	
	\While {$i \leq K$ $\cap$ $v \geq \text{SQV}_i^{n}$}
	\State  $i \leftarrow$ $i + 1$
	\EndWhile 
	
	$r_p \leftarrow 2(\frac{i}{K} - 0.5)$
   
\EndIf

\If {$a^n(t) == a^n(t-1)$}
	\State $r_p \leftarrow c_1 \cdot r_p $
    \Comment{$c_1$ is constant $>$ 1}

\EndIf
\end{algorithmic}
\end{algorithm}

The second component of the reward mechanism is designated as the \emph{social welfare reward}, denoted by $r_{sw}$. This reward takes into consideration the implications of a network's actions on the individual rewards of its neighboring networks during the interval between two consecutive decision-making instances. Specifically, a network is categorized as a neighboring network if the Euclidean Distance (ED) between their network center points is less than a predetermined threshold, represented as $\Gamma$ meters. In essence, $r_{sw}$ captures the collective welfare and interdependence among the networks, accounting for the influence exerted by the behavior of each network on the rewards of others over successive iterations. The social welfare reward is calculated as the arithmetic mean of the personal rewards of all neighboring networks. The pseudo-code for the social welfare reward is outlined in Algorithm \ref{alg:algo3}.

\begin{algorithm}
\caption{CARLTON's Social Welfare Reward, $r_{sw}$}
\label{alg:algo3}
\begin{algorithmic}[1]
\INPUT $R_p$, $i$, $\mu$, $\Gamma$  \Comment{List of all $r_p$ of other networks between two consecutive decision points, the index of the network of interest, dictionary of all center points, and constant value}
\OUTPUT $r_{sw}$  

\State $n \leftarrow 0 $ 
\State $r_{sw} \leftarrow 0$
\For {$j \in \{1,...,N\}$}
	\If {ED($\mu[j]$,$\mu[i]$) $\leq \Gamma $ $\cap$ $j \neq i$} 
		\State $n \leftarrow n + 1$
		\State $r_{sw} \leftarrow \frac{1}{n} \left((n-1)r_{sw} + r_p^j\right)$
		
	\EndIf
\EndFor

\end{algorithmic}
\end{algorithm}

Finally, the total reward is a linear combination of the two reward terms:
\begin{equation}\label{reward_G}
r = \rho r_p + (1-\rho) r_{sw},
\end{equation}
where $\rho$ $\in [0,1]$ is a design parameter.

\subsection{Training Procedure}\label{train}

CARLTON adheres to the multi-agent paradigm known as Centralized Training with Decentralized Execution (CTDE). At the beginning of each episode, each individual agent, denoted as the network manager, is assigned an exclusive serial number. This serial number determines the timing of its actions, effectively simulating a delay time characteristic of real-world scenarios. This technique is strategically employed to circumvent the potential collision of decisions among different networks and to mitigate the overall dynamic complexity of the scenario. To ensure a consistent and balanced decision-making process, the scenario horizon is explicitly defined as $T$ times the total number of participating networks within the given scenario. Consequently, this arrangement allows each network to possess precisely $T$ decision points per scenario, guaranteeing equitable opportunities for each participant to influence the outcome of the scenario.

The resource allocation concerning channel frequency selection during the training phase conducted by each agent adheres to the following prescribed methodology. In particular, with a probability denoted by $\epsilon_b$, the agents draw samples from the distribution in (\ref{dis2}), originally introduced in the Exp3 algorithm for the non-stochastic multi-armed bandit problem \cite{auer2002nonstochastic}. Otherwise, the agents follow a greedy policy:
\begin{equation}
\label{dis2}
\begin{array}{l}
\displaystyle
\text{Pr}(a^n(t) = k| s^n(t))\vspace{0.2cm}\\
\displaystyle
\frac{(1-\alpha)e^{\beta Q(s^n(t),k;\theta)}}{\sum_{a' \in \{1,...K\}} e^{\beta Q(s^n(t), a';  \theta)}} + \frac{\alpha}{K} \;\; \forall \ k\in\mathcal{K}.
\end{array}
\end{equation}
Here, $s^{n}(t)$ stands for the state at time step $t$ of network $\mathcal{N}^n$, $Q(s^n(t),k;\theta)$ stands for the estimated action value by the NN with weights $\theta$, which its architecture described below, and $\alpha$ and $\beta$ are constants. Additionally, during the training period comprising $B$ episodes, we annealed $\epsilon_b$ from 0.5 to 0.01 after each of the first $B/2$ episodes, and then maintained it at a steady value for the remaining half, i.e., the latter $B/2$ episodes.

Furthermore, an additional step involves the application of an action-masking technique over the $Q(s^n(t), a^{n}(t))$ values, which is based on the $\text{QV}^{n}$ values on each channel $k$. For any channel in $\text{QV}^{n}$ with a value of $0$, the probability of selecting this channel is masked to $0$, while in the event of $\text{QV}^{n} == \overrightarrow{0}$, no action selection is performed, and the previous action is preserved. This technique serves the purpose of preventing collisions among different networks and is mathematically executed as follows:
\begin{equation}\label{maskingEq}
    Q(s^{n}(t), k; \theta)= 
\begin{cases}
    -\infty ,& \text{if} \ \text{QV}^{n}_k = 0, \\
    Q(s^{n}(t), k;\theta) , & \text{otherwise.}
\end{cases}
\end{equation}
 
Next, as is inherent in DRL value-based algorithms, our principal objective is to find an estimating function, denoted as $f_{\theta}$, which approximates the action value $Q(s^{n}(t), a^{n}(t);\theta)$. This function adeptly maps any given set of state and action $(s, a)$ to a real number within the domain of $\mathbb{R}$. The optimization of $f_\theta$ is commonly achieved through gradient descent algorithms over a differentiable loss function. In this context, we employ the Huber loss to compute our loss value. The Huber loss function is given by:
\begin{equation}\label{huberloss}
    L_{\delta}(e)= 
\begin{cases}
    \frac{1}{2} e^2 ,& \text{if} \  |e|\leq \delta,\\
    \delta \cdot (|e| - \frac{1}{2} \delta) ,              & \text{otherwise,}
\end{cases}
\end{equation}
where  $e^{n} = (r^{n}(t) + \gamma \underset{a'\in \mathcal{K}}{mm_\omega}Q(s^{n}(t+1),a';\theta) - Q(s^{n}(t),a^{n}(t);\theta))$, and $mm_{\omega}$ is a smooth maximum operator named Mellowmax operator with a temperature parameter $\omega$ \cite{asadi2017alternative}. Generally, $e$ is the error between the target and predicted values, and $\delta$ is a discrimination hyper-parameter threshold. Subsequently, with the aid of these estimated action values, the learn policy can be derived. To enable this process, we formally define the input state as follows:
\begin{equation}
s^{n}(t) = \text{concatenate}(\text{CBR}, \text{QV}^{n}(t)).
\end{equation}
Here, the term $\text{CBR}$ represents the Channel Binary Representation, refers to the channel that the network utilizes before executing the action. In other words, it is a binary representation of the preceding action if the action was fulfilled. The term $\text{QV}^{n}(t)$ stands for the $\text{QV}^{n}$ at time  step $t$.

The NN architecture of CARLTON consists of three dense layers with skip connections. The hidden layer uses the Leaky-ReLU activation function with a slope of $0.2$, while the output layer employs the identity activation function. Furthermore, to initialize the weights of the NN, the Glorot-Uniform method, as introduced in the work by Glorot and Bengio \cite{glorot2010understanding}, is applied, and the biases are initialized to zero.

The training procedure unfolds across $B$ scenarios as follows: First, within each main training loop, we uniformly sample the number of participating networks. Subsequently, we generate a scenario following the rules outlined in Section \ref{environment}. For each network, we then randomly assign a network number $n\in{1,...,N}$ to define $\mathcal{N}^n$ and allocate a unique replay memory ($\mathcal{D}^{n}$). Next, we iterate over $T$ time steps for each scenario, where at each step we execute actions for each network sequentially, corresponding to their assigned network number. Then, we store the relevant information in the respective unique replay buffer. Upon scenario completion, we transfer data from each $\mathcal{D}^{n}$ into a global replay memory (GRM). Subsequently, we uniformly sample from the GRM and train the NN by minimizing (\ref{huberloss}) $N_E$ times.

The training procedure is illustrated in Algorithm \ref{alg:algo4}, providing a comprehensive outline of the steps involved. For a clear overview of the hyperparameters utilized during the training process, please refer to Table \ref{table:hyperparameters}.

\begin{algorithm}[!h]
\caption{CARLTON's Training Procedure}
\label{alg:algo4}
\begin{algorithmic}[1]
\INPUT All hyperparameters in Table \ref{table:hyperparameters}
\INPUT Initialize NN weights, $\theta_0$ 
\OUTPUT Trained mechanism for distributed DCA, $\theta^*$
\State {Create Global Replay memory ($\text{GRM}$) of size $Sz$ }
\For {$i \in  \{1,...,B\}$}
    \State {Uniformly sample random number of networks}
	\State {Set a random scenario with $N$ networks (Section \ref{environment})}
        \State {Randomly match unique network numbers}
	\State {Create Local Replay Memory ($\mathcal{D}^{n}$) for each network }
	\State {Randomly initialize each network to some channel}

	\For {$t \in \{1,...,T\}$}
        \For{$n \in \{1,...,N\}$}
    		\State obtain $\text{QV}^{n}(t)$
            \State  $s^{n}(t) \leftarrow $concatenate(\text{BCR},$\text{QV}^{n}(t)$) 
    		\State Store ($s^{n}(t-1)$, $a^{n}(t-1)$, $r^{n}(t)$, $s^{n}(t)$) in $\mathcal{D}^{n}$
    		\State {Select $a^{n}(t)$ for $\mathcal{N}^{n}$ based on $f_{\theta_{i-1}}$}
    		 \State Execute $a^{n}(t)$ and observe $r^{n}_p$
        \EndFor
	\EndFor
	 \State Store all data in local $\mathcal{D}^{n}$ in $\text{GRM}$ $\forall n\in\{1,...,N\}$ 
	\For {$\{1,...,N_{E}\}$}
	\State 	  Uniformly sample a mini-batch of size $\text{bz}$,
		\State  $(s(t),a(t),r(t),s(t+1)) \sim \text{GRM}$ 
		 \State Train the NN by minimizing (\ref{huberloss}) and obtain $\theta_i$	
	\EndFor 
\EndFor
\end{algorithmic}
\end{algorithm}

\begin{table}[!h]
\centering
\begin{tabular}{|m{5cm}  m{3cm} |} 
 \hline
 Parameter & Value  \\ [1ex] 
  
 \hline\hline 
 Number of channels, $K$ & $10$\\
 \hline 
 Channels frequencies carriers, $\{f_1,..., f_k\}$ & $\{208, ..., 226| $\\
 $[MHz]$	& $f_{k+1} - f_{k} = 2 MHz\}$ \\
 
 \hline
 Antennas gains, $G_R, G_T$  & $1$\\
 \hline
Antennas heights ($H_R, H_T$) [m] &  $1$\\

\hline 
Transmit power, $\text{PT}$ [dBW] & $2$\\
\hline 
 Range of initial center point, $u_1$ $[m]$ & $400$\\
 \hline
 Radius range, $x_1, x_2$ $[m]$ & $50,500$\\
 \hline 
 The set of possible users per network, $\mathcal{M}$ & $\{1,...,15\}$\\
 
 \hline 
 SINR threshold, $\text{SINR}^* [dBm]$  & 4 \\
 \hline 
  Desired reward, $r_{desired}$ & 4\\
  \hline 
   Constant, $c_1$ & 1.1 \\
 \hline
 A desired threshold, $\zeta$ & 0.9 \\
 \hline 
 Euclidean distance threshold which defines a network as a neighbor $\Gamma$ [m] & $500$ \\ 
 \hline 
 Personal reward fraction, $\rho$ & 0.7\\
 
 \hline 
 Distribution parameters $\alpha$, $\beta$ & 0, 1\\
 
  \hline 
 Huber Loss parameter, $\delta$ & 1\\
 
 \hline
 Number of episodes during training, $B$ & $1000$    \\  
  
  \hline 
 MellowMax parameter, $\omega$ & 0.02, if $i\leq N/2$, \\

 & 0.2, otherwise \\ 
 
  \hline 
 Discount factor, $\gamma$ & 0.9 \\
  \hline 
 Optimizer & Adam \\
 
  \hline 
 Optimizer learning rate &  0.00025, if $i\leq N/2$, \\
  & 0.0001, otherwise \\
  
  \hline 
 Optimizer exponential decay rate for 1st, and 2nd moment estimates, $\beta_1$, $\beta_2$ & 0.9, 0.999 \\
   
 \hline 
 Optimizer epsilon for numerical stability, $\epsilon_{Adam}$ & $10^{-7}$\\
   
 \hline
  Global Replay Memory size, $S_z$ & $10^5$\\
 \hline
  Number of decision points per network, $T$ (also the size of the Local replay Memory) & 20 \\

 \hline
 NN architecture &    \\
 Number of hidden layer & $3$  \\
 Nodes at each layer & $128$    \\
 Number of skip connection (addition after activation) & 2 \\
 Hidden layer activation function & $\text{Leaky-ReLU}$ slop of 0.2\\
 Output activation function & Identity \\
 Weights initialization technique & Glorot-Uniform\\
 Bias initialization & $\vec{0}$\\

 \hline
 Number of training steps, $N_E$ & $40$    \\ 
  
 \hline

 Batch size, $bz$ & $32$   \\ [1ex] 
 \hline
\end{tabular}
\caption{A description of the hyper-parameter values used in the experiments conducted in this research.}
\label{table:hyperparameters}
\end{table}

\section{Experiments and Discussion}

In this section, we present comprehensive experimental results to illustrate the efficiency of the proposed CARLTON algorithm in a real-world simulation environment, as described in Section \ref{sec:system}. We commence by analyzing specific features in the training procedure of CARLTON in Subsection \ref{sim:carlton}. Subsequently, we provide a performance comparison with state-of-the-art methods in Subsection \ref{sim:comparison}. 

\subsection{CARLTON's Training Results}
\label{sim:carlton}

\begin{figure}[ht!]
\vspace{-0.0cm}
\begin{center}
\advance\leftskip-3cm
\advance\rightskip-3cm

\includegraphics[width=0.50 \textwidth,trim={0.cm 0.0cm 1.0cm 1.0cm},clip ]{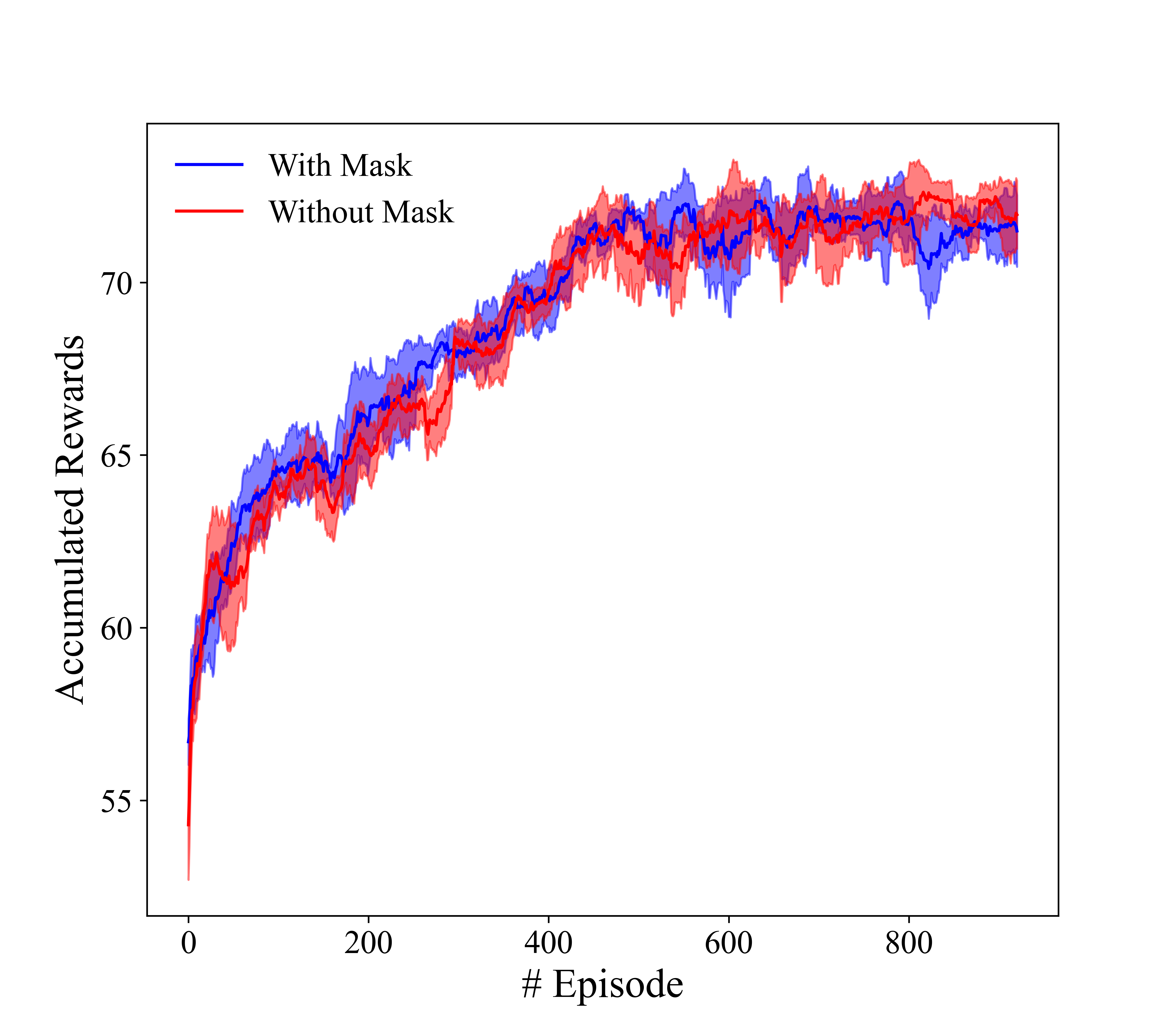} 

\caption[]{CARLTON performance: The accumulated rewards as function of episode number during training for implementation of with and without masking approach. In a case of perfect game the maximum value is 88.}
\label{fig:3}
\end{center}
\end{figure}

CARLTON undergoes two distinct training scenarios, one involving the action masking approach, as referenced in (\ref{maskingEq}), and the other without it. As expected, the incorporation of masking during action selection results in higher accumulated rewards during the initial phases of training (Episodes $\leq 400$). This improvement stems from the reduction of unnecessary exploration, as illustrated in \Figref{3}. However, as training progresses and convergence is achieved (Episodes $\geq 500$), both techniques yield comparable results. It is important to note that both approaches remain applicable when the messages exchanged between network users and the network manager are transmitted through a control channel, which is assigned and freed for communication at any given time.

\begin{figure}[ht!]
\vspace{-0.0cm}
\begin{center}
\advance\leftskip-3cm
\advance\rightskip-3cm
\includegraphics[width=0.50 \textwidth,trim={0.cm 0.0cm 1.0cm 1.0cm},clip ]{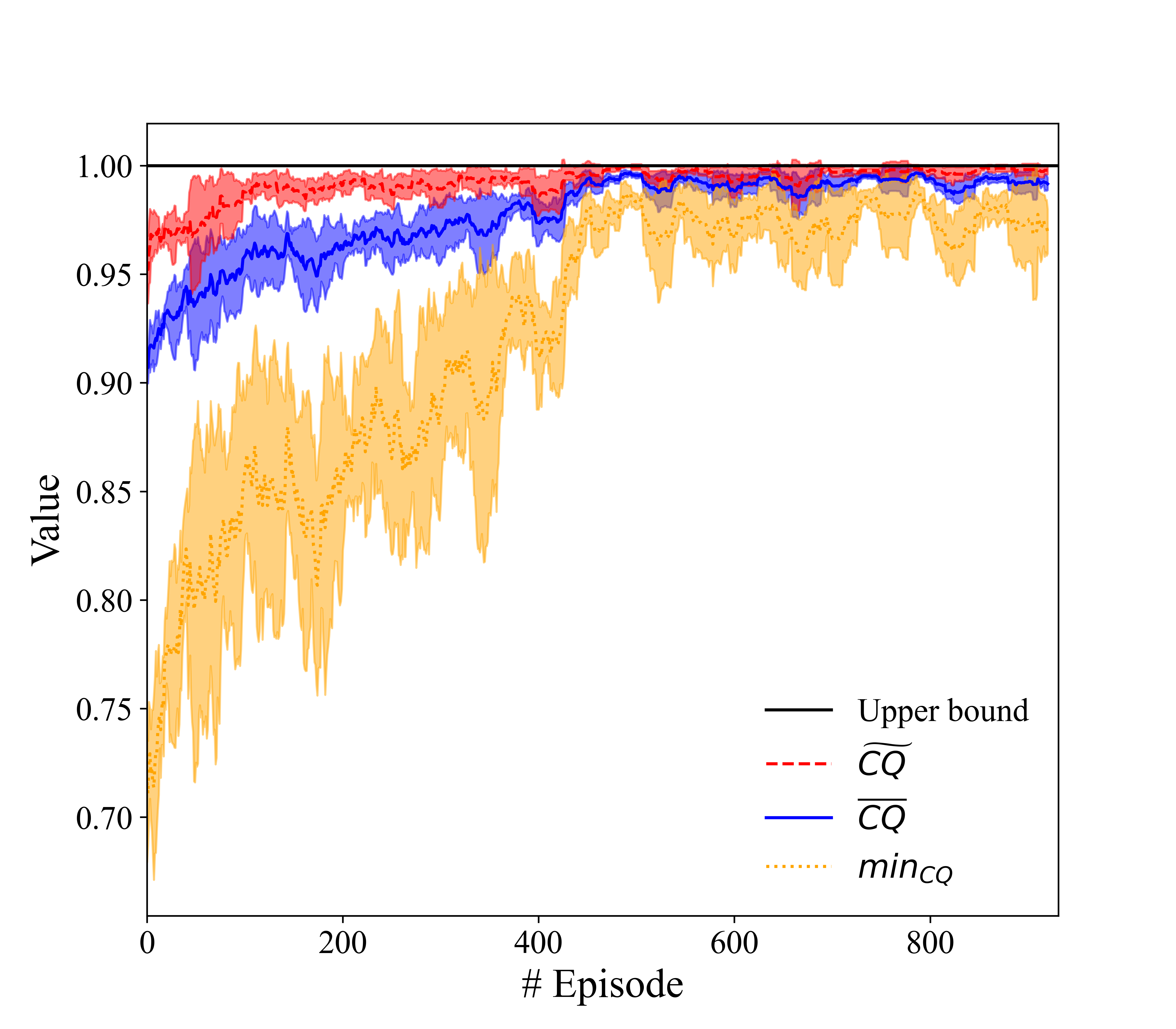} 
\caption[]{CARLTON performance: The empirical results of the mean ($\overline{CQ}$), median ($\widetilde{CQ}$), minimum ($min_{CQ}$) values of channel quality during training.}
\label{fig:4}
\end{center}
\end{figure}

On the other hand, in situations where no dedicated control channel is available, the utilization of masking becomes essential to ensure the existence of a functional communication link between the users and the network manager. Thus, it guarantees that there is at least some form of connection facilitating communication and data exchange. In addition, using the masking approach not only depends on the presence or absence of a control channel for communication but also helps to fast convergence in case of higher action space, as shown in previous work \cite{vinyals2017starcraft,berner2019dota,ye2020mastering,hou2023exploring}. When directing our attention to the physical results obtained during the training process, it is evident that CARLTON exhibits noteworthy enhancements in the mean, median, and minimum values of the channel quality. The improvement of these values, as defined in (\ref{params1}), are shown in  \Figref{4}. Furthermore, upon convergence, it is conspicuous that the agents have acquired the ability to cooperate effectively, resulting in a scenario where the lowest channel quality ($min_{CQ}$) of some agents in the system remains above $0.95$ percent. Here, $c^{n}(t)$ is the channel of network $n$ at time step $t$.

\begin{equation}\label{params1}
\begin{gathered}
\text{CQ} = \left[\text{QV}^{1}_{c^1(T)}(T), ...,\text{QV}^{N}_{c^N(T)}(T)\right],\vspace{0.2cm}\\
\overline{CQ} = \mathbb{E}[\text{CQ}],\vspace{0.2cm} \\ 
\widetilde{CQ} = \text{Median}(\text{CQ}),\vspace{0.2cm}\\
min_{CQ} = min(\text{CQ}).
\end{gathered}
\end{equation}

Similarly, we have investigated the progress observed during training regarding supplementary physical parameters. The first additional parameter is the average number of channel changes (ANCC), defined as the average number of channel changes until convergence per network. The ANCC score (ANCCS) is given by:
\begin{equation}
\text{ANCCS} = 1 - \frac{\text{ANCC}}{T}.
\end{equation}

The second parameter is the convergence time (CT), which refers to the time of the last channel change made by one of the networks during the scenario. The convergence time score (CTS) is calculated as follows:
\begin{equation}
\text{CTS} = 1 - \frac{\text{CT}}{T\cdot N}\;.
\end{equation}

The third parameter, denoted as spectrum efficiency (SE), is intended to assess the spectral distortion resulting from the utilization of networks after convergence or at the end of the episode. A higher value of SE parameter indicates that the integration of a new network into the group becomes an easier task by providing more channels with high quality to operate on, along with causing minimal distortion to the convergences solution achieved by the earlier participants. The SE score (SES) is computed as follows:

\begin{equation}
\begin{gathered}
\text{SES} = \mathbb{E}\left[\Psi\right] = \frac{1}{N}\sum_{n=1}^{N}\Psi^{n},\vspace{0.2cm}\\
\hspace{-7.5cm}\text{where}\vspace{0.2cm}\\
\Psi^{n} = \left(\sum_{k=1}^{K}\left(\text{QV}^{n}_k(T)\right)^{2}\right)^{0.5}/{K}^{0.5}.
\end{gathered}
\end{equation}

The last parameter that we define is a linear weighted score (WS) of the parameters of interest, given by: 
\begin{equation}
\text{WS} = w_1 \cdot \overline{\text{CQ}} + w_2 \cdot \text{ANCCS} + w_3 \cdot \text{\text{CTS}} +w_4 \cdot \text{SES},
\end{equation} 
where we set $w_1$, $w_2$, $w_3$, $w_4$ in the simulations to $0.4, 0.1, 0.4, 0.1$, respectively.

\begin{figure}[ht!]
\vspace{-0.0cm}
\begin{center}
\advance\leftskip-3cm
\advance\rightskip-3cm

\includegraphics[width=0.50 \textwidth,trim={0.cm 0.0cm 1.0cm 1.0cm},clip ]{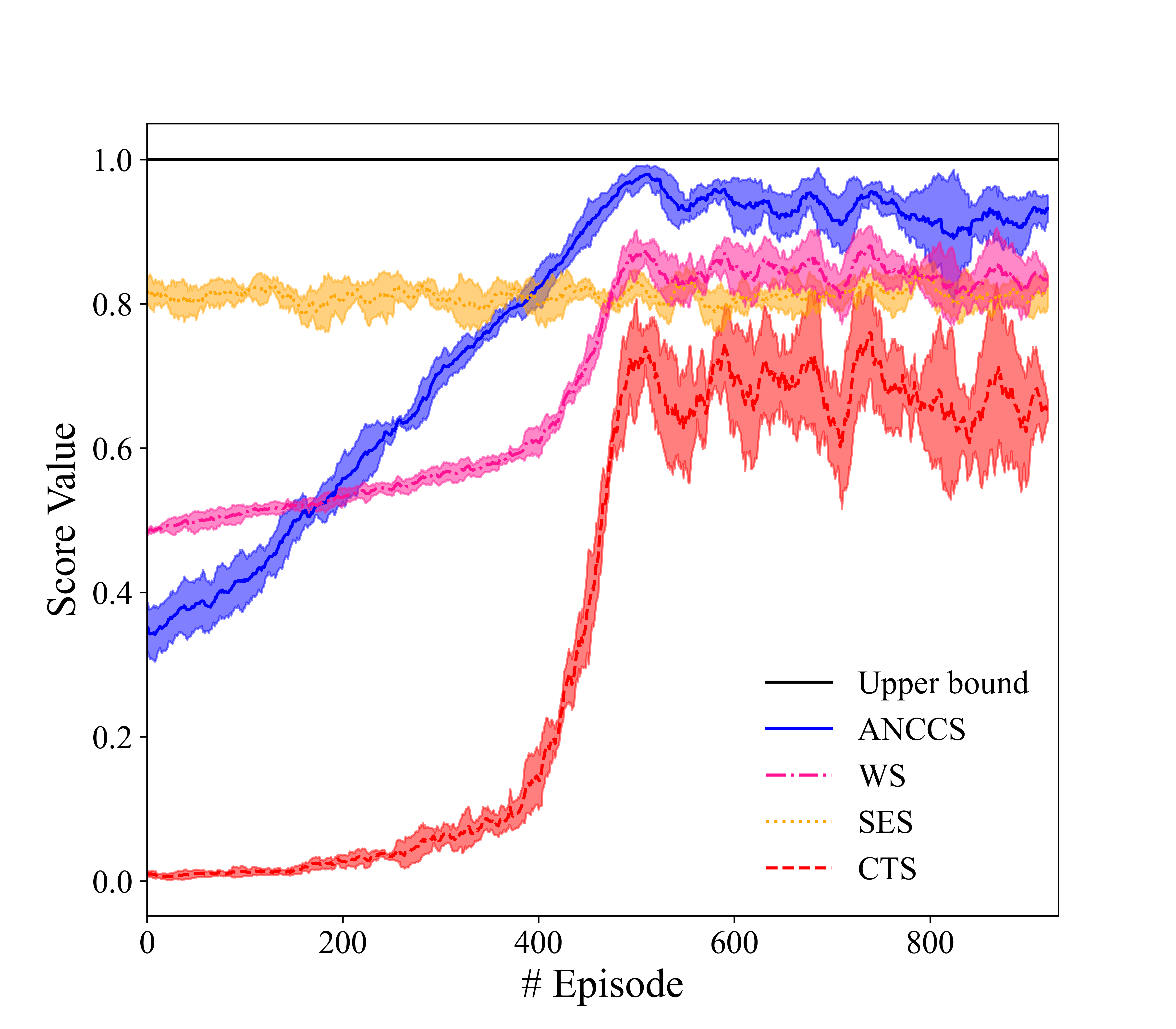} 

\caption[]{CARLTON performance: The score values of the average number of channel changes (ANCC), convergence time (CT), spectrum efficiency (SE), and the weighted score (WS) during training.}
\label{fig:5}
\end{center}
\end{figure}

As apparent from \Figref{5}, all parameter scores, except SE, demonstrate improvement. These results align with our reward function, as specified in (\ref{reward_G}), wherein the impact of the CT and ANCC on the reward value is given in line $10$ in Algorithm \ref{alg:algo2}. Moreover, in view of the absence of a direct motivation to enhance the SE score, the algorithm possesses the freedom to establish an SE score that harmonizes with the converged policy, and results in an average value of approximately 0.8 for the SE score.

\begin{figure}[ht!]
\vspace{-0.0cm}
\begin{center}
\advance\leftskip-3cm
\advance\rightskip-3cm

\includegraphics[width=0.50 \textwidth,trim={0.cm 0.0cm 1.0cm 1.0cm},clip ]{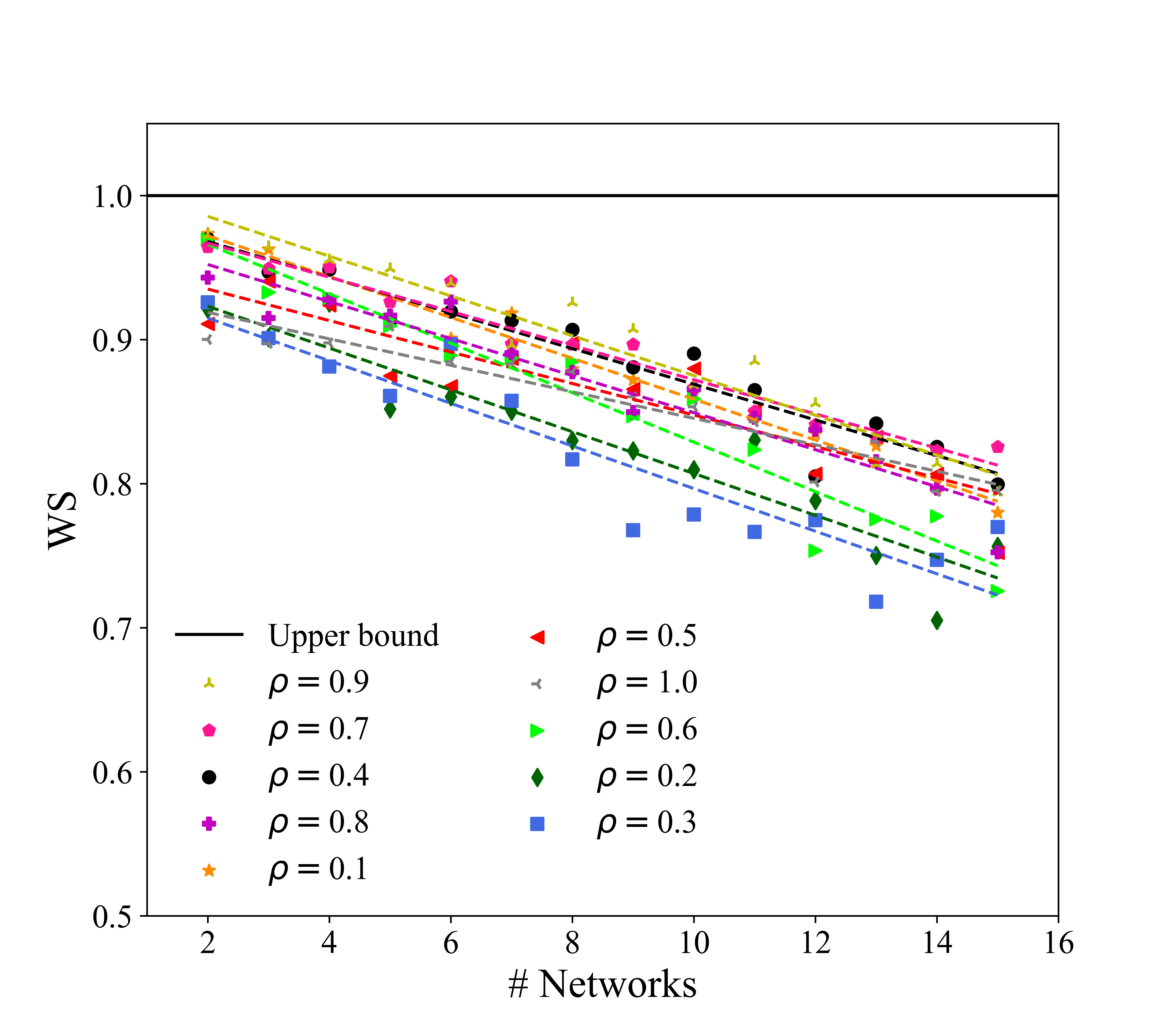} 

\caption[]{CARLTON performance: The weighted score value, $\text{WS}$, as a function of the number of networks participating in the scenario for different values of $\rho$.}
\label{fig:6}
\end{center}

\end{figure}

\begin{figure}[ht!]
\vspace{-0.0cm}
\begin{center}
\advance\leftskip-3cm
\advance\rightskip-3cm

\includegraphics[width=0.50 \textwidth,trim={0.cm 0.0cm 1.0cm 1.0cm},clip ]{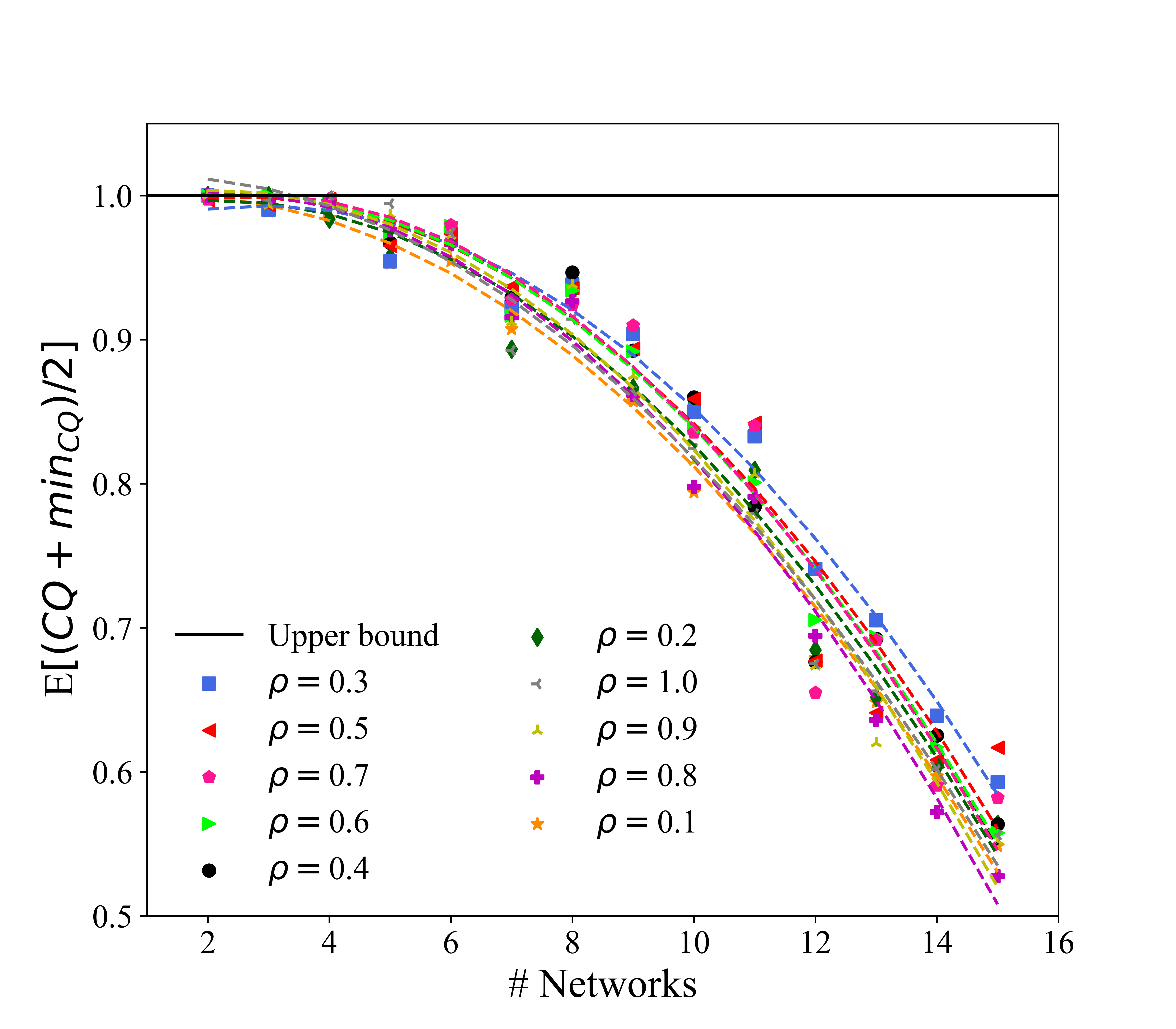} 

\caption[]{CARLTON performance: the expectation of the mean value of $\text{CQ}$ and $min_{QC}$ as a function of the number of networks participating in the scenario for different values of $\rho$.}
\label{fig:7}
\end{center}

\end{figure}

Moreover, we conducted an analysis of the influence of the personal reward weight, denoted as $\rho$ in (\ref{reward_G}), on the algorithm's physical behavior. All options were tested over 30 games for each scenario, where the scenarios differed by the number of networks participating, ranging from 2 to 15 networks. In total, this amounted to 420 games. Despite the training process being limited to scenarios involving a maximum of 7 networks, our examination extends to include the evaluation of the algorithm's generalization capabilities across scenarios with a larger number of participating networks. It was observed that, with respect to the metric $\text{WS}$ as depicted in \Figref{6}, higher values demonstrated superior performance and enhanced generality. Conversely, in the context of $\text{CQ}$ and $min_{CQ}$, lower values exhibited better performance, as evidenced by the expectation of the mean value displayed in \Figref{7}. This disparity in results can be attributed to the fact that higher values lead to more aggressive policies, resulting in quicker convergence. On the other hand, lower values facilitated increased cooperation and fairness among the participating networks, leading to a slower but more stable mechanism. A value of 0.7 was found to effectively strike a balance between these two contrasting aspects, rendering it an effective selection. In the context of cooperative missions, this equilibrium point is particularly noteworthy, as it arises from the operational constraints of decentralized agents without information exchange capabilities, effectively achieving equilibrium between self-interest and social responsibility.
  
\begin{figure}[ht!]
\vspace{-0.0cm}
\begin{center}
\advance\leftskip-3cm
\advance\rightskip-3cm

\includegraphics[width=0.50 \textwidth,trim={0.cm 0.0cm 1.0cm 1.0cm},clip ]{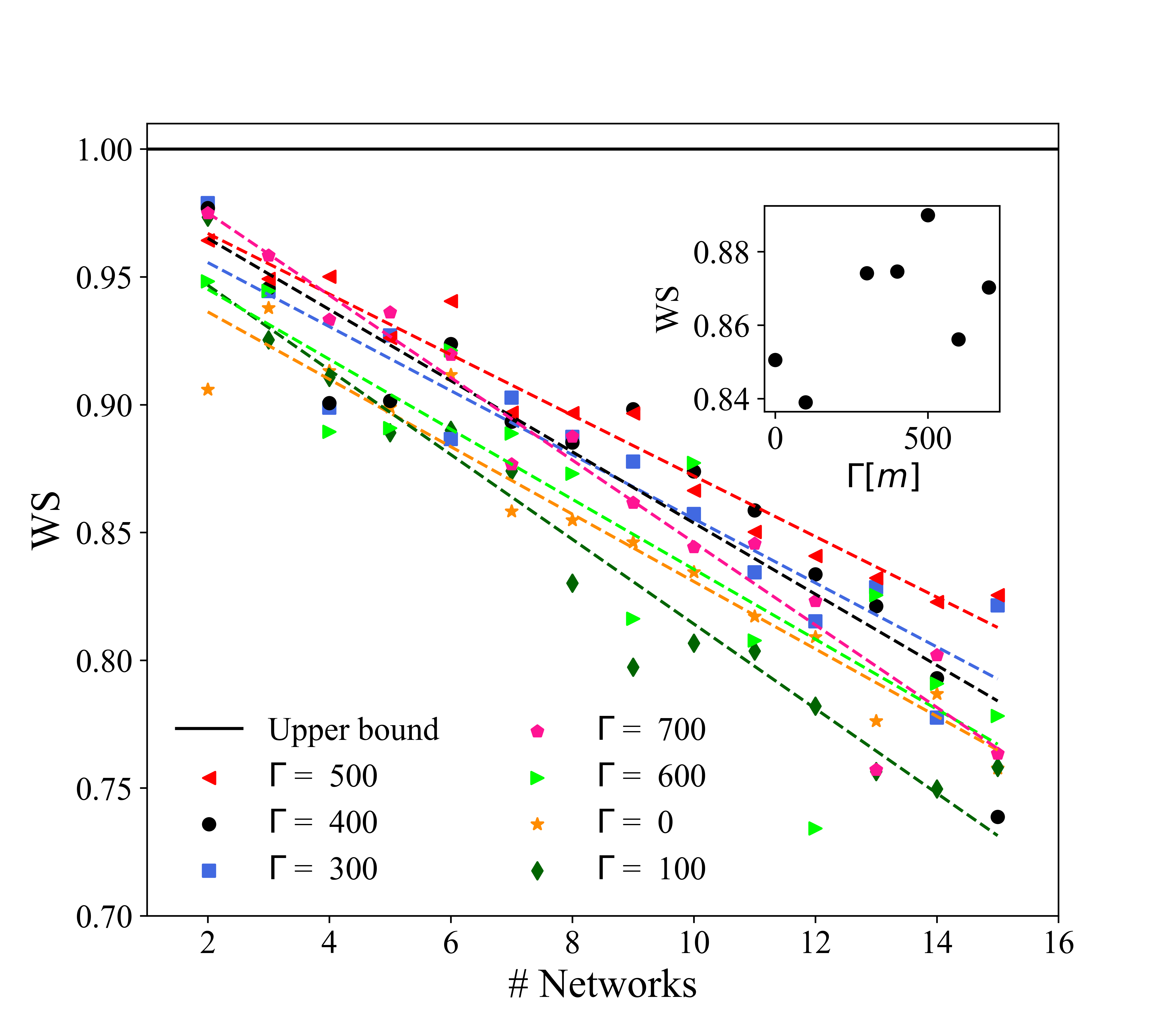} 

\caption[]{CARLTON performance: The weighted score value, $\text{WS}$, as a function of the number of networks participating in the scenario for different values function of $\Gamma$ [m]. Inset: The average value of $\text{WS}$ across all scenarios as function of $\Gamma$ [m].}
\label{fig:72}
\end{center}

\end{figure}

Afterward, we conducted an analysis pertaining to the influence of a designated threshold parameter denoted as $\Gamma$, which serves to establish the criteria for considering networks as neighbors. To briefly recapitulate, if the ED between the central points of two networks falls below the threshold $\Gamma$, both networks are deemed neighbors. These neighbor networks influence each other's rewards through the parameter denoted as $r_{sw}$, as shown in (\ref{reward_G}) and elaborated in Algorithm \ref{alg:algo3}. In the scenario where $\Gamma$ approaches zero ($\Gamma \rightarrow 0 \ [m]$), each network progressively exhibits more self-centered behavior, prioritizing its individual performance exclusively. Conversely, when $\Gamma$ tends towards infinity ($\Gamma \rightarrow \infty$), each network gradually adopts a more socialistic stance, with its reward encompassing a broader spectrum of network performances, including those with which it shares no direct impact on their $\text{QV}$s. For example, when the Euclidean Distance (ED) between the central points equal to $700 [m]$, there exists a $3\%$ probability that the two networks will exert an influence on each other's $\text{QV}$s. Our analysis unveiled that, concerning WS performance, designating a network as a neighbor when the distance falls below $500$ meters (corresponding to a probability of $50\%$ that both networks influence each other's $\text{QV}$s) within the environmental framework described in Section \ref{environment}), yielded the highest average performance across all tested scenarios, as depicted in \Figref{72}.

\begin{figure}[ht!]
\vspace{-0.0cm}
\begin{center}
\advance\leftskip-3cm
\advance\rightskip-3cm

\includegraphics[width=0.50 \textwidth,trim={0.cm 0.0cm 1.0cm 1.0cm},clip ]{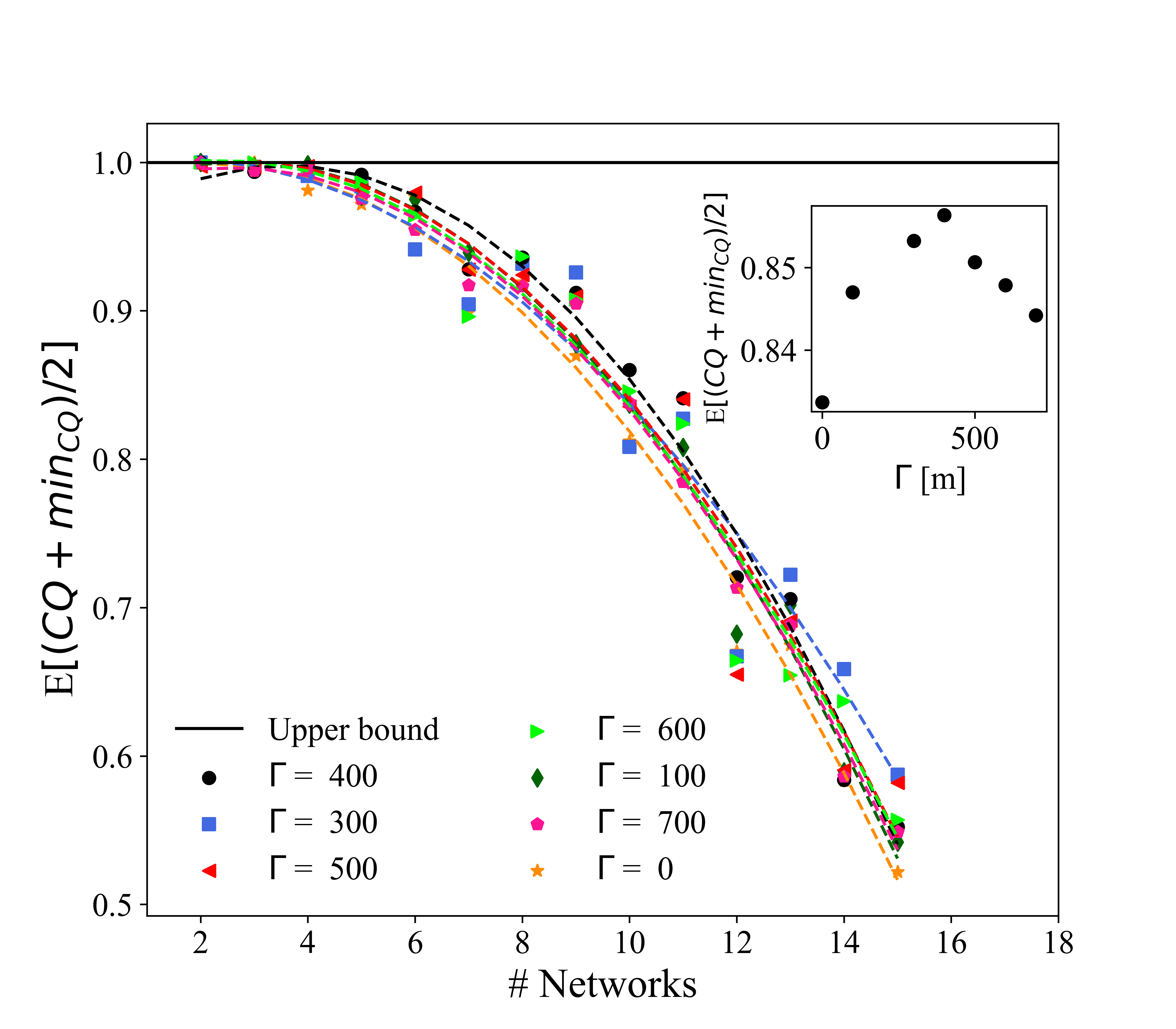} 

\caption[]{CARLTON performance: The expectation of the mean value of $\text{CQ}$ and $min_{QC}$ as a function of the number of networks participating in the scenario for different values of $\Gamma$ [m]. Inset: The average value of $E[(\text{CQ}+min_{CQ})/2]$ across all scenarios as function of $\Gamma$ [m].}
\label{fig:73}
\end{center}

\end{figure}

On the contrary, when we shift our attention to the expectation value of $[(CQ+min_{CQ})/2]$, it becomes evident that $\Gamma = 400$ meters (representing a probability of $86\%$ that the networks impact each other's $\text{QV}$s) yields the most favorable average value across the entire domain under examination, as illustrated in Figure \ref{fig:73}. These findings underscore the notion that, in order to achieve a high level of social performance encompassing all participants within the given scenario, a network should only consider interactions with those networks that have a high probability of interaction. This concept not only benefits individual performance but also augments the performance of the entire system. Conversely, it is evident that adopting and learning a selfish policy ($\Gamma = 0$) exhibits the least favorable performance across all scenarios examined. It is noteworthy that when taking into account the average of both parameters, $\Gamma = 500$ meters stands out as the leading choice.

Subsequently, we sought to further enhance the performance of our proposed algorithm by implementing a post-processing step in the decision-making mechanism. Given the algorithm's generalization across diverse and extensive domains involving varying numbers of participating networks, certain instances of unnecessary channel switches were observed. For example, in a 2-network scenario, the algorithm would switch from a channel with a $\text{CQ}$ of 1.0 to another channel with the same $\text{CQ}$ value, even when this switching had no impact on the other participant. Such occurrences are a result of the algorithm's generalization mechanism, which has learned to create sparsity and adopt policies that aim to provide favorable performance in numerous situations. To address this issue, we introduced a post-processing step as follows: If the $\text{CQ}$ of the desired channel is not greater than the $\text{CQ}$ of the current channel by at least $\varphi \%$, the algorithm maintains its current channel, and effectively abstained from unnecessary changes. However, if the difference in $\text{CQ}$ values exceeds the specified threshold, the algorithm follows the proposed action and executes the channel switch accordingly. The baseline algorithm, without post-processing, is denoted by $\varphi = $ None. Furthermore, in multi-agent systems with a high number of members, it can be challenging to account for every possible scenario. Consequently, the agents tend to learn policies that exhibit bias toward the scenarios within the training set. Applying a threshold based on channel quality can reduce unnecessary spectrum mobility, which serves as a post-processing block. To illustrate, consider a scenario in which a single network, initially operating on a randomly selected channel, might transition to an alternative channel without any rational reason, a behavior likely stemming from the biased policy learned by the agents. 

\begin{figure}[ht!]
\vspace{-0.0cm}
\begin{center}
\advance\leftskip-3cm
\advance\rightskip-3cm

\includegraphics[width=0.50 \textwidth,trim={0.cm 0.0cm 1.0cm 1.0cm},clip ]{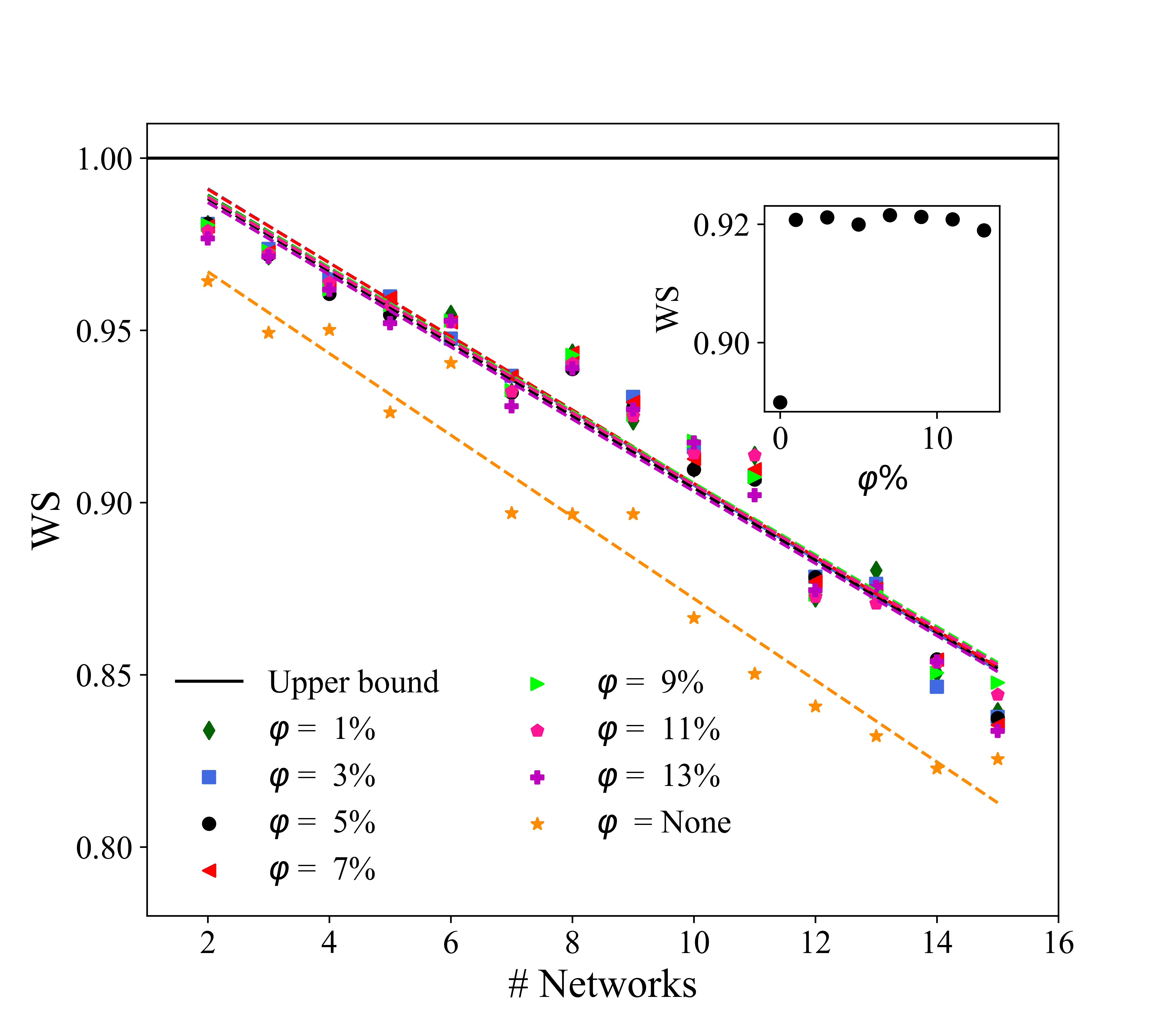} 

\caption[]{CARLTON performance: The expectation of the weighted score, $\text{WS}$, as a function of participating network at each scenario. Inset: The average value of $\text{WS}$ across all scenarios as function of the threshold value $\varphi$, 0 stands for None.}
\label{fig:8}
\end{center}

\end{figure}

\begin{figure}[ht!]
\vspace{-0.0cm}
\begin{center}
\advance\leftskip-3cm
\advance\rightskip-3cm

\includegraphics[width=0.50 \textwidth,trim={0.cm 0.0cm 1.0cm 1.0cm},clip ]{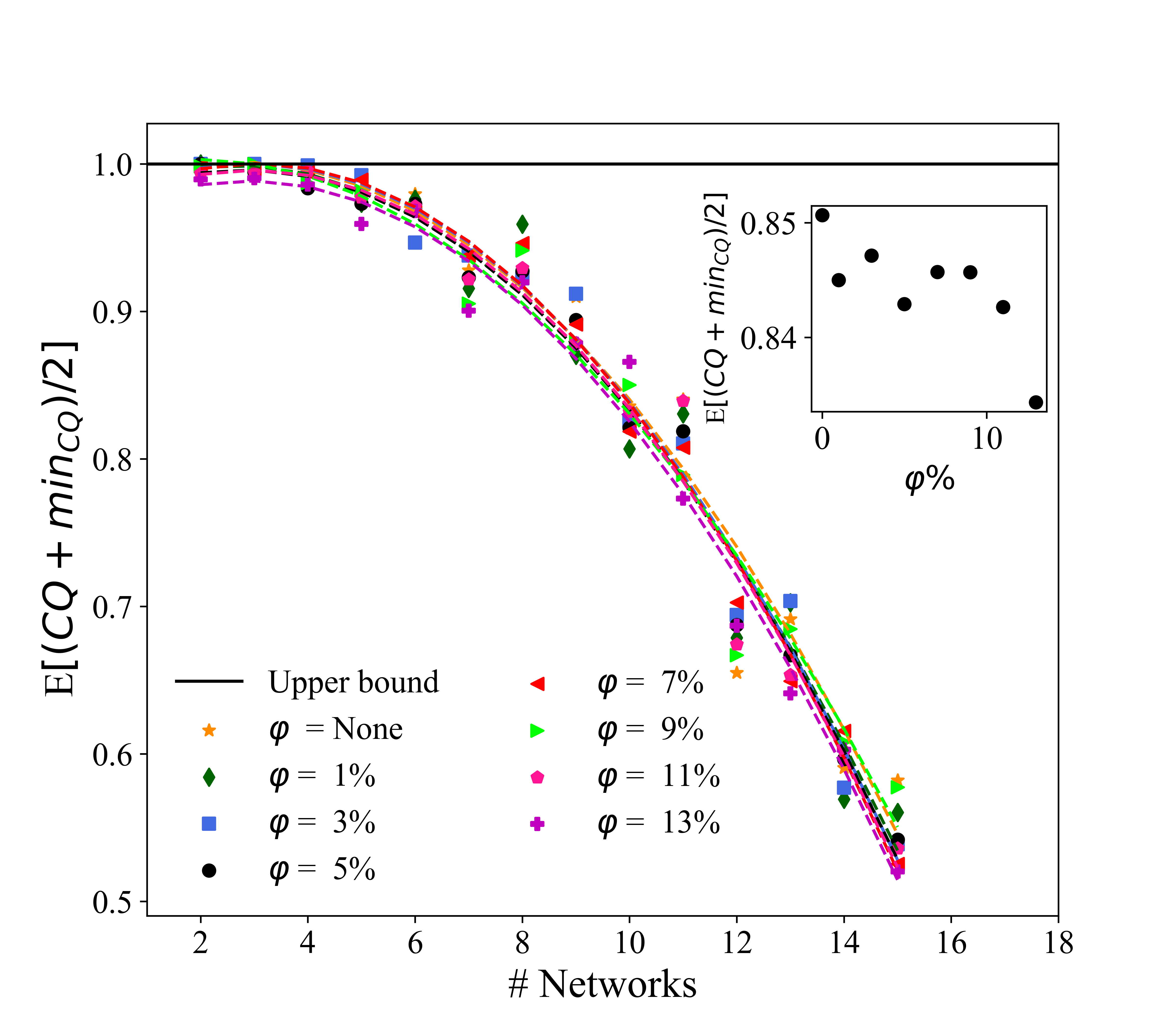} 

\caption[]{CARLTON performance: The expectation of the average between $\text{CQ}$ and $min_{QC}$ as a function of participating network at each scenario. Inset: The average value of $E[(\text{CQ}+min_{CQ})/2]$ across all scenarios as function of the threshold value $\varphi$, 0 stands for None.}
\label{fig:9}
\end{center}

\end{figure}

The findings illustrate a correlation between the increase in the weighted score as the threshold value, $\varphi$, increases, as evident from the observations in \Figref{8}. Conversely, there is a decrease in the channel quality, as represented in \Figref{9}. This indicates that the inclusion of the threshold significantly expedites the convergence time, albeit at the cost of achieving a better solution concerning channel quality. The selection of the optimal threshold value is contingent upon two crucial factors, the size of the network, where the switch to another channel can be expensive in terms of communication operations, and the quality threshold set by the user. These considerations emphasize the significance of setting a thoughtful threshold to achieve a trade-off between convergence speed and channel quality. The introduction of a post-decision mechanism, as an optional modification, holds promise for overall performance enhancement, thereby warranting further investigation and experimentation to assess its potential benefits fully.

\subsection{Performance Comparison} 
\label{sim:comparison}

We now present a comprehensive comparative analysis to assess the performance of CARLTON compared to other well-known algorithms. To establish an upper bound, we considered an algorithm based on graph coloring, which operates in a centralized manner and is not suitable for distributed implementations. Furthermore, we conducted a comprehensive evaluation of CARLTON against alternative approaches, including Jammer Avoidance Response (JAR) \cite{huang2022prospects}. In the case of JAR, designed to mitigate interference or jamming, the algorithm draws inspiration from electric fish mobility, aiming to identify potential jamming and autonomously shift its electric discharge frequency away from the potential jamming frequency. JAR executes channel switches to nearby channels ($\pm 2MHz$) only if such switches can enhance the channel quality by a margin of at least 0.05. This approach has demonstrated state-of-the-art results within the considered model compared to existing methods. Additionally, we evaluated against the well-known Random Agent (RA) \cite{cohen2017distributed} static initialization, where we randomly set the initial channel and refrain from making any further channel switches. The comparison with these alternative algorithms provided valuable insights into the effectiveness and advantages of CARLTON within the considered model. 

\begin{figure}[ht!]
\vspace{-0.0cm}
\begin{center}
\advance\leftskip-3cm
\advance\rightskip-3cm

\includegraphics[width=0.50 \textwidth,trim={0.cm 0.0cm 1.0cm 1.0cm},clip ]{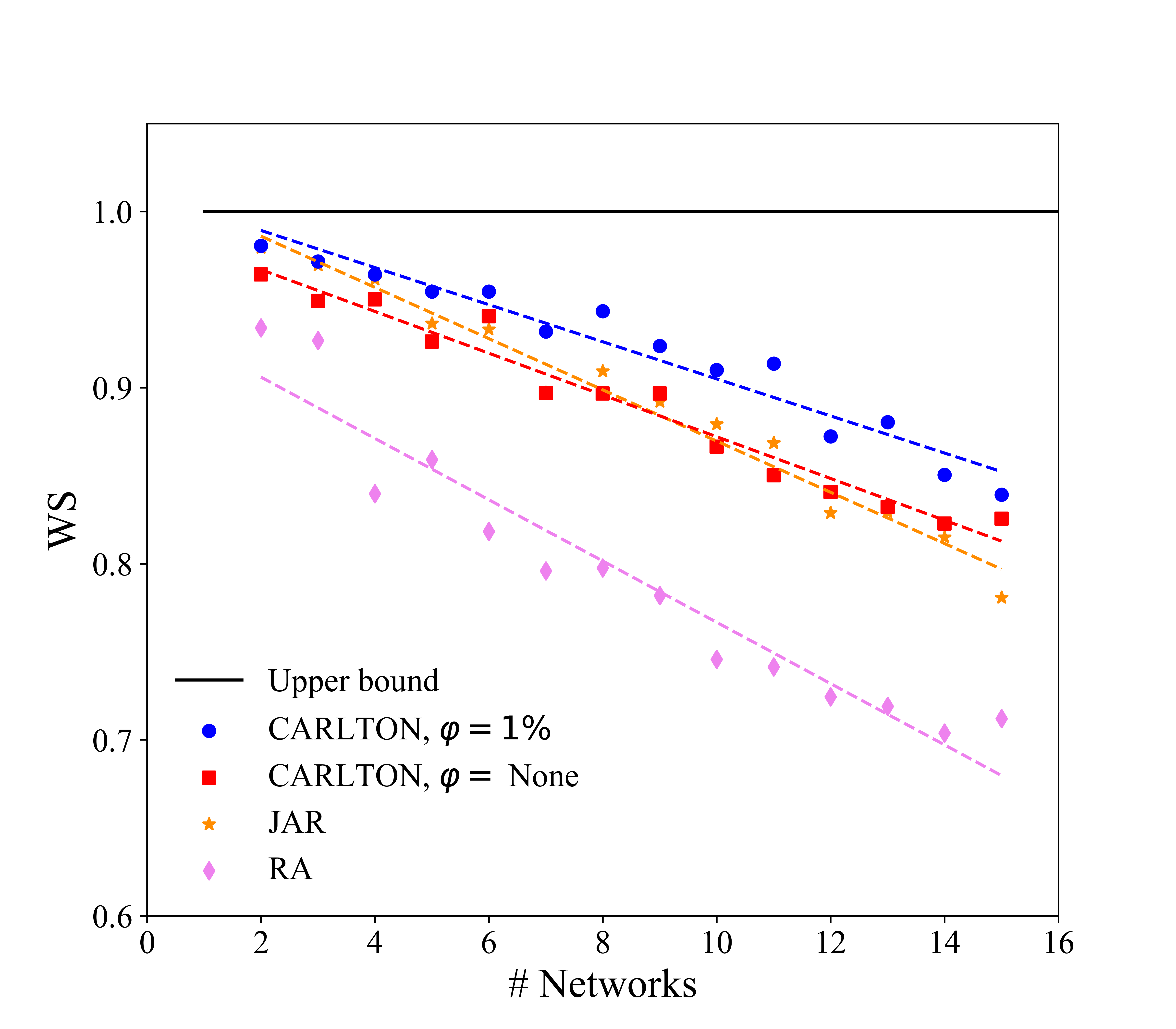} 

\caption[]{Algorithm comparison: The weighted score value, $\text{WS}$, as a function of the number of networks participating in the scenario.}
\label{fig:10}
\end{center}

\end{figure}

\begin{figure}[ht!]
\vspace{-0.0cm}
\begin{center}
\advance\leftskip-3cm
\advance\rightskip-3cm

\includegraphics[width=0.50 \textwidth,trim={0.cm 0.0cm 1.0cm 1.0cm},clip ]{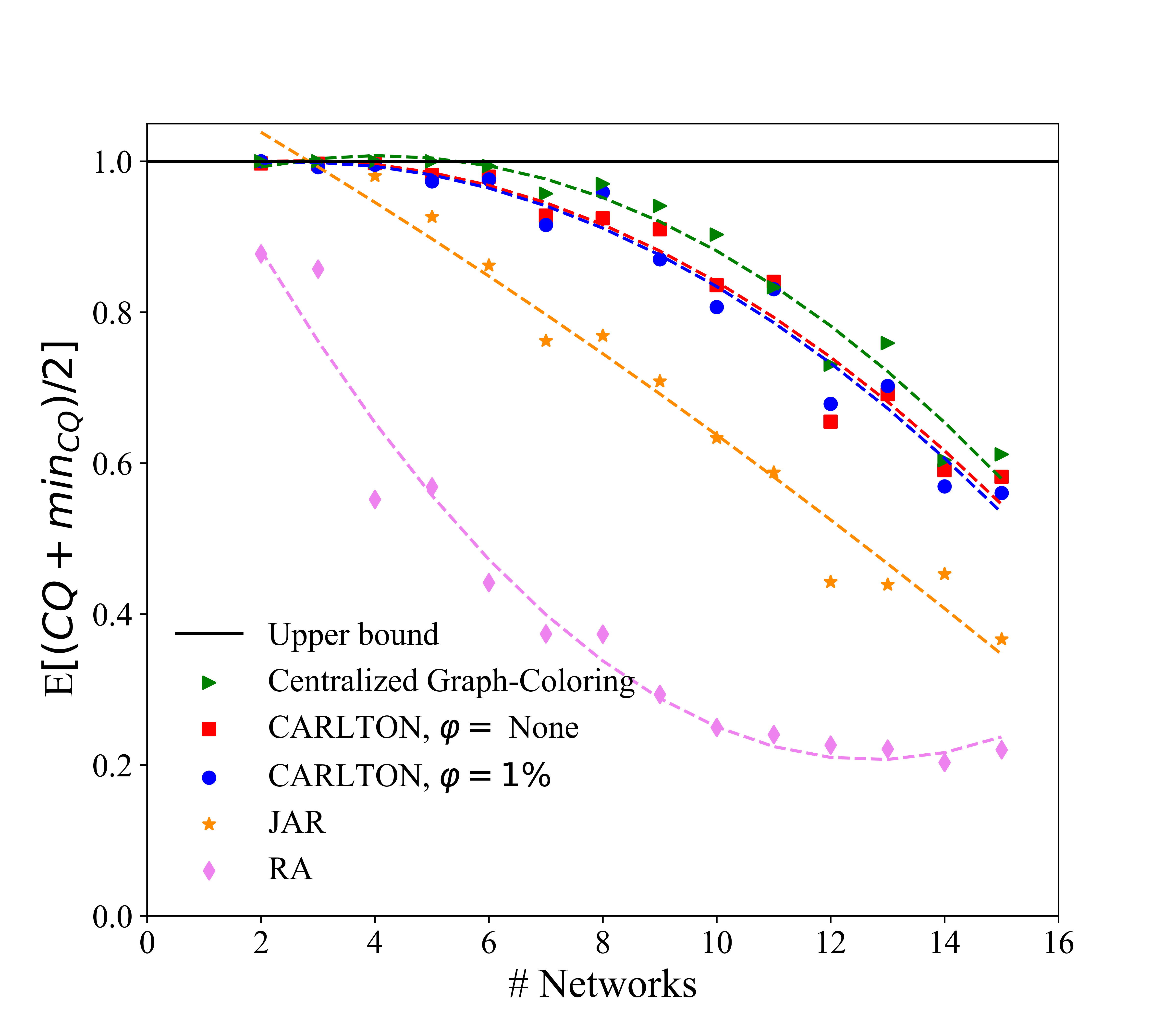} 

\caption[]{Algorithm comparison: The expectation of the mean value of $\text{CQ}$ and $min_{QC}$ as a function of the number of networks participating in the scenario.}
\label{fig:11}
\end{center}

\end{figure}

In a manner analogous to the comparison involving $\rho$, all algorithms underwent testing across the identical set of 420 games. The obtained results demonstrated the superior performance of CARLTON in terms of both the $\text{WS}$ metric and the expectation of $[\text{CQ} + min_{CQ}]/2$ values, as illustrated in \Figref{10} and \Figref{11}, respectively.

\begin{figure}[ht!]
\vspace{-0.0cm}
\begin{center}
\advance\leftskip-3cm
\advance\rightskip-3cm

\includegraphics[width=0.50 \textwidth,trim={0.cm 0.0cm 1.0cm 1.0cm},clip ]{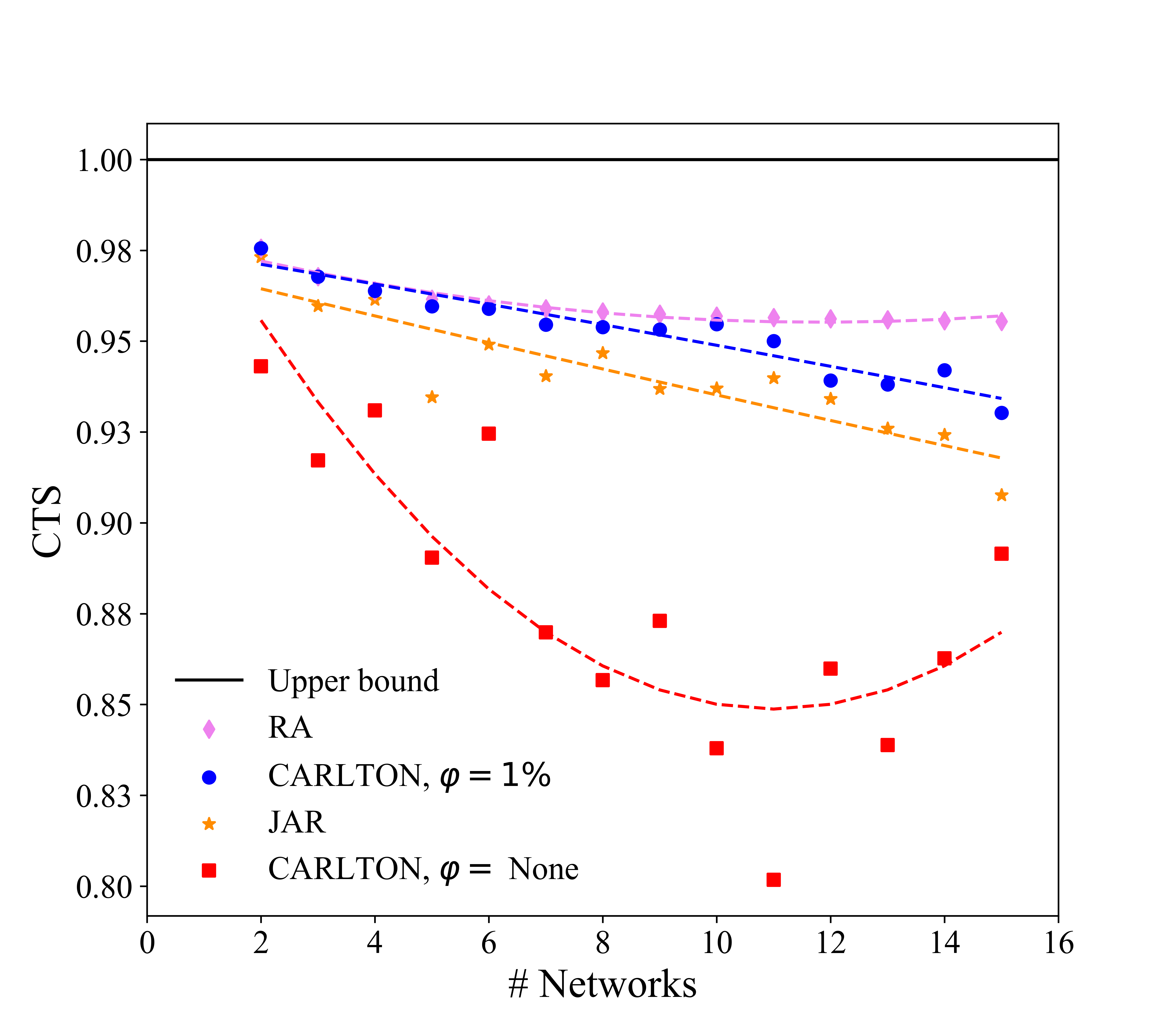} 

\caption[]{Algorithm comparison: The convergence time score (CTS) as a function of the number of networks participating in the scenario.}
\label{fig:11b}
\end{center}

\end{figure}

\begin{figure}[ht!]
\vspace{-0.0cm}
\begin{center}
\advance\leftskip-3cm
\advance\rightskip-3cm

\includegraphics[width=0.50 \textwidth,trim={0.cm 0.0cm 1.0cm 1.0cm},clip ]{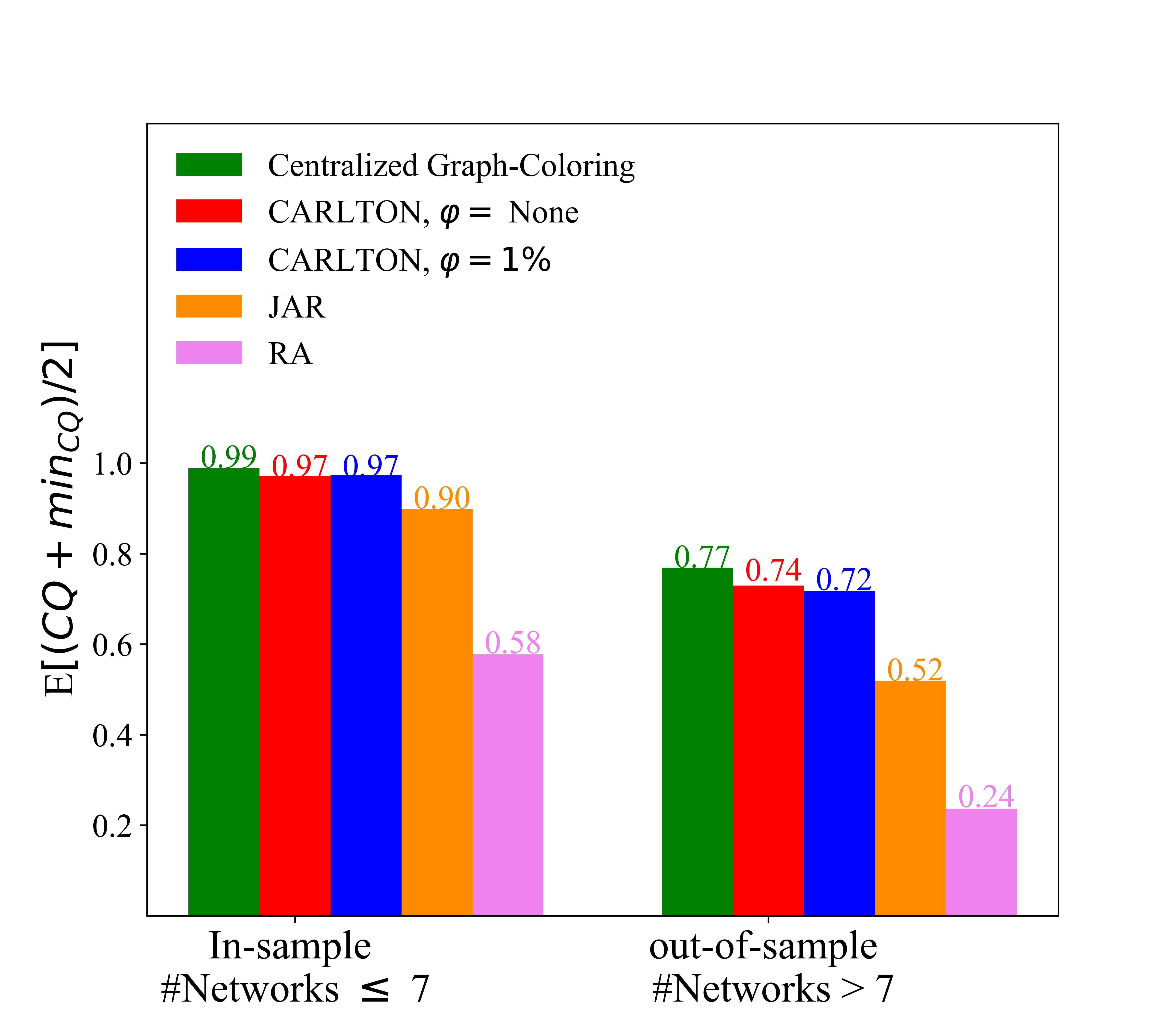} 

\caption[]{Algorithm comparison: The average value of $E[(\text{CQ}+min_{CQ})/2]$ across all scenarios for each algorithm. For CARLTON $\varphi = 0$.}
\label{fig:12}
\end{center}

\end{figure}

Nevertheless, as shown in \Figref{11b}, the convergence time score demonstrates lower values in the case of CARLTON without post-processing ($\varphi = None$), indicating that achieving superior performance in channel quality required a more comprehensive interaction among the network participants than alternative algorithms, translating into a longer time. Furthermore, CARLTON displayed a marginal difference of only $2\%$ compared to the centralized graph-coloring approach in the in-sample domain (i.e., $\#Networks<7$), underscoring its efficacy in distributed systems, as illustrated in \Figref{12}.

Finally, the obtained results unequivocally demonstrate the exceptional performance of CARLTON, even in scenarios where it was not explicitly trained on (i.e., $\#Networks > 7$, named out-of-sample), thereby highlighting its remarkable generality. This remarkable adaptability can be attributed to the underlying concept that scenarios with a high number of networks in certain constellations can be effectively represented by combinations of multiple scenarios involving a lower number of networks. Consequently, the agent exhibits cooperation not with all networks but rather selectively with a subset, likely comprising fewer than 7 networks, which are considered its neighboring networks. This premise elucidates the compelling rationale behind the observed excellent generality.\\

\section{Conclusion}\label{sec:conclusion}

We have introduced a novel distributed DCA algorithm, dubbed CARLTON, designed for multi-agent systems within the framework of MARL. CARLTON is specifically tailored for scenarios involving non-orthogonal operating channels with ICI, necessitating cooperation among participating networks to enhance spectral efficiency. The key emphasis of CARLTON lies in networks with a substantial number of users, aiming to achieve both high channel quality and rapid convergence. Our experimental results highlight the superior performance of CARLTON compared to alternative approaches, exhibiting only a marginal trade-off in channel quality when compared to the centralized-based graph-coloring approach. Moreover, CARLTON demonstrates remarkable generalization capabilities, effectively handling scenarios that were not encountered during training (out-of-sample scenarios). This effectiveness positions CARLTON as an exceptional algorithm for proficiently managing DCA among distributed cognitive networks in complex interference environments.

\setcounter{figure}{0}                       
\renewcommand\thefigure{A.\arabic{figure}} 
\setcounter{algorithm}{0}
 \renewcommand\thealgorithm{A.\arabic{algorithm}} 

\bibliographystyle{ieeetr}%
\bibliography{ref} 

\end{document}

%% file: macros.tex
\newcommand{\beq}{\begin{equation}}
\newcommand{\eeq}{\end{equation}}
\newcommand{\bea}{\begin{array}}
\newcommand{\ena}{\end{array}}
\newcommand{\bds}{\begin {itemize}}
\newcommand{\eds}{\end {itemize}}
\newcommand{\bdf}{\begin{definition}}
\newcommand{\blm}{\begin{lemma}}
\newcommand{\edf}{\end{definition}}
\newcommand{\elm}{\end{lemma}}
\newcommand{\bthm}{\begin{theorem}}
\newcommand{\ethm}{\end{theorem}}
\newcommand{\bprp}{\begin{prop}}
\newcommand{\eprp}{\end{prop}}
\newcommand{\bcl}{\begin{claim}}
\newcommand{\ecl}{\end{claim}}
\newcommand{\bcr}{\begin{coro}}
\newcommand{\ecr}{\end{coro}}
\newcommand{\bquest}{\begin{question}}
\newcommand{\equest}{\end{question}}

\newcommand{\larrow}{{\larrow}}

\def\urltilda{\kern -.15em\lower .7ex\hbox{\~{}}\kern .04em}